\documentclass{article}

\usepackage{arxiv}
\usepackage{xcolor}
\usepackage[utf8]{inputenc}
\usepackage[T1]{fontenc}
\usepackage{hyperref}
\usepackage{url}
\usepackage{booktabs}
\usepackage{amsfonts}
\usepackage{microtype}
\usepackage{graphicx}
\usepackage{natbib}
\usepackage{doi}
\usepackage{amsmath,mathtools,amssymb}
\usepackage{dsfont}
\usepackage{float}

\newcommand{\bphi}{\boldsymbol\phi}

\newcommand{\RD}{\mathrm{RD}}
\newcommand{\LRR}{\mathrm{logRR}}
\newcommand{\LOR}{\mathrm{logOR}}
\newcommand{\LHR}{\mathrm{logHR}}
\newcommand{\Var}{\operatorname{Var}}
\newcommand{\Eff}{\operatorname{Eff}}

\newcommand{\Cov}{\operatorname{Cov}}

\newcommand{\Se}{\mathrm{Se}}
\newcommand{\Sp}{\mathrm{Sp}}

\title{From Test Performance to Risk-Based Effect Sizes: A Unified Wald-Type Framework to Design Clinical Validation Studies for Binary and Survival Outcomes}

\date{}

\author{
    {Yongqi Zhong} \\
    Adela Inc.\\
	Foster City, CA \\
	{yongqi.zhong@adelabio.com} \\
    \And
    {Anne-Renee Hartman} \\
    Adela Inc.\\
	Foster City, CA \\
	{anne-renee.hartman@adelabio.com} \\
	\And
	{Jing Zhang} \\
    Adela Inc.\\
	Foster City, CA \\
	{jing.zhang@adelabio.com} \\
}

\renewcommand{\headeright}{}
\renewcommand{\undertitle}{}
\renewcommand{\shorttitle}{Efficient Design and Sample Size for Predictive Test Validation}

\begin{document}
\maketitle

\begin{center}
\begin{minipage}{0.92\linewidth}
\small
\noindent\textbf{Co-corresponding authors:} Yongqi Zhong (yongqi.zhong@adelabio.com) and Jing Zhang (jing.zhang@adelabio.com), Adela Inc., Foster City, CA.

\noindent\textbf{Conflicts of Interest:} Y.\ Zhong and J.\ Zhang report full-time employment with Adela, Inc., and Adela stock options. A.-R.\ Hartman reports full-time employment in a leadership role at Adela, Inc., and stock/stock options in Adela, Inc., Delfi Diagnostics, and Mirvie.

\noindent\textbf{Funding:} This study was sponsored by Adela, Inc..

\noindent\textbf{Acknowledgments:} The authors thank Melanie Styers (Medical and Scientific Communications, Adela, Inc.) for editorial assistance in the preparation of this manuscript.

\end{minipage}
\newline
\newline
\end{center}

\begin{abstract}
Clinical validation studies of predictive tests are usually designed to focus on sensitivity (Se) and specificity (Sp), while statistical power is often calculated on regression-effect scales (e.g., risk ratio, hazard ratio). However, these quantities are statistically connected. Here, we provide closed-form links from sensitivity, specificity, and disease prevalence ($\pi$) to predictive risks, risk contrasts, and Wald-type variance, power, and sample-size formulas for binary and fixed-horizon survival outcomes. Analyses of statistical efficiency via C- and D-optimal principles demonstrate how prevalence and threshold choices affect study efficiency, supporting rapid decisions in preliminary studies and informing the design of subsequent, larger studies. Simulations show good calibration across most realistic scenarios; when events are rare and test effects are simultaneously very large, continuity and minimum-event corrections are needed to stabilize the approximation. We illustrate the framework with a case study describing use of the coronary artery calcium score for predicting incident cardiovascular disease in patients with type 2 diabetes mellitus. The formulas let investigators check power and required enrollment directly from $(\Se,\Sp,\pi)$, without running a separate simulation for each design candidate.
\end{abstract}


\section{Introduction}
\label{sec:intro}

Evaluation of predictive tests and models has become routine in clinical research, particularly in cardiovascular risk stratification, biomarker validation, and other longitudinal risk-prediction settings. Validation studies typically ask a practical question: does a test result predict a patient's risk with sufficient accuracy to inform clinical decision-making \cite{steyerberg_assessing_2010,vickers_decision_2006}? Sensitivity (Se) and specificity (Sp) quantify performance of the test based on discrimination at a given threshold, but clinical decisions are generally informed by predicted risks and contrasts on risk, odds, or hazard scales \cite{walcott_measuring_2021,bowling_methods_2023}.

Guidance for clinical validation study design is divided between two related literatures. Classifier accuracy research emphasizes Se/Sp and predictive values \cite{pepe_limitations_2004,pepe_integrating_2008,brenner_variation_1997}, while the effect-size and power literature usually starts from regression coefficients in generalized linear or proportional-hazards models \cite{shieh_power_2005,hsieh_sample-size_2000,schoenfeld_asymptotic_1981,chow_sample_2017,riley_evaluation_2024}. Existing sample-size calculation formulas for predictive values stop at the precision of predictive values rather than risk-contrast estimands built from both predictive strata \cite{steinberg_sample_2009}. Recent threshold evaluation sample-size work has the same gap \cite{whittle_extended_2025}. As a result, investigators often cannot determine whether a design specified by sensitivity, specificity and disease prevalence has enough information for clinically targeted contrasts on risk, odds, or hazard scales.

The challenge in integrating sensitivity/specificity with risk-based outcomes is typified by studies leveraging the coronary artery calcium score (CACS) in cardiovascular medicine. CACS is derived from cardiac computed tomography and is widely used to stratify future cardiovascular disease (CVD) risk, particularly in asymptomatic individuals and in higher-risk groups, such as patients with type 2 diabetes mellitus (T2DM) \cite{koo_coronary_2023}. In this setting, a binary classification (e.g., CACS at or above a chosen Agatston cutoff vs.\ below) is obtained at baseline, and patients are followed for incident CVD, where inference may target either fixed-horizon risk contrasts (e.g., risk difference) or time-to-event contrasts (e.g., hazard ratio) \cite{koo_coronary_2023}. Similar design problems in clinical validation studies arise for other biomarkers and clinical prediction models used in longitudinal care \cite{us_food_and_drug_administration_hematologic_2020,us_centers_for_medicare__medicaid_services_lcd_nodate}.

Investigators often have anticipated sensitivity and specificity from development or pilot data, but power calculations are still run with regression-based formulas that assume direct effect-size parameterization. Hence, a single design framework that connects classifier performance characteristics to effect size is needed.

\paragraph{Contributions and structure.}
We derive a Fisher information matrix on the stratum risk parameters $(p_1,p_0)$ with test-positive fraction $\tau$---all obtained by composition from $(\Se,\Sp,\pi)$---then read off closed-form variance, power, and required $N$ for any smooth contrast $g(p_1,p_0)$, including RD, logRR, logOR, and fixed-horizon logHR. The design workflow used throughout the paper is
\begin{align*}
&\underbrace{(\Se,\Sp,\pi)}_{\text{test performance assumptions}} 
\;\longrightarrow\;
\underbrace{(\mathrm{PPV},\mathrm{NPV})}_{\text{predictive values}}
\;\longrightarrow\;
\underbrace{(p_1,p_0)}_{\text{stratum risks}}
\;\longrightarrow\;
\underbrace{\Delta}_{\text{effect size}}
\\
&\qquad\longrightarrow\;
\underbrace{\mathcal{I}_{\psi}}_{\text{Fisher information}}
\;\longrightarrow\;
\underbrace{\Var(\widehat{\Delta}),\ N}_{\text{precision, sample size}}
\;\longrightarrow\;
\text{Design Efficiency}.
\end{align*}
Section~\ref{sec:variance_framework} develops the risk and effect-size parameterization. Section~\ref{sec:wald} gives Fisher-information variance and Wald-type sample size and power formulas. Section~\ref{sec:design_optimality} uses optimal design theory (e.g., C- and D-optimal criteria) \cite{pukelsheim_optimal_2006} for efficient pilot-design decisions, and Section~\ref{sec:casestudy} applies the framework to coronary artery calcium scoring for cardiovascular disease risk validation.
For practical use, Algorithm~1 summarizes the design workflow from $(\Se,\Sp,\pi)$ to $(p_1,p_0,\tau)$, the effect estimand $\Delta=g(p_1,p_0)$, and required sample size.
\begin{center}
\fbox{%
\parbox{0.97\linewidth}{%
\textbf{Algorithm 1: Design from test accuracy inputs}
\begin{enumerate}
\item Specify test accuracy inputs $(\Se,\Sp)$ and prevalence $\pi$ (optionally specify plausible ranges for sensitivity analysis).
\item Compute the test-positive fraction $\tau=\Pr(X=1)$ and implied outcome risks $p_1=\Pr(Y=1\mid X=1)$ and $p_0=\Pr(Y=1\mid X=0)$ from $(\Se,\Sp,\pi)$ under the joint $(X,Y)$ model.
\item Choose the clinical effect estimand $\Delta=g(p_1,p_0)$ to compare strata (e.g., RD, logRR, logOR; and for survival, a fixed-horizon complementary log-log contrast at $t^\ast$, logHR), and select the link $f(\cdot)$ used for Wald testing.
\item Obtain $\Var(\widehat{\Delta})$ from the information-based/delta-method variance derived below for $\Delta$ (using the model for $(p_1,p_0)$ and the gradient of $g$), and compute required sample size $N$ from the Wald normal-approximation power formula at the desired $\alpha$ and power.
\item Apply the recommended small-sample safeguards (e.g., continuity correction and minimum expected events checks). Sensitivity across plausible $(\Se,\Sp,\pi)$ scenarios can then be evaluated if needed.
\end{enumerate}
}%
}
\end{center}

\section{From Test Performance Metrics to Risk-based Effect Sizes for Binary and Survival Outcomes}
\label{sec:variance_framework}
\subsection{From Test Accuracy to Predictive Utility}

Let $Y\in\{0,1\}$ denote target status and $X\in\{0,1\}$ the test result. We define

\begin{equation*}
    \Se=\Pr(X=1\mid Y=1),\qquad
    \Sp=\Pr(X=0\mid Y=0),\qquad
    \pi=\Pr(Y=1).
\end{equation*}
Here $\pi$ denotes prevalence in the intended-use population for the validation study.

Sensitivity and specificity condition on $Y$ and describe discrimination at a fixed threshold. Positive and negative predictive values condition on $X$ and therefore quantify clinically interpreted risks after observing the test result \cite{bowling_methods_2023,brenner_variation_1997}. Because development cohorts may differ from deployment populations and design inputs $(\Se,\Sp,\pi)$ are target-population quantities, these values taken directly from model-development cohorts can be optimistic or miscalibrated for clinical validation studies \cite{steyerberg_assessing_2010,ben-haim_interpreting_2024,riley_evaluation_2024}.

\subsection{From Test Performance to Conditional Risks}
\label{sec:from-metrics-to-effect}
The joint distribution of $(X,Y)$ has multinomial cell probabilities
\begin{equation*}
    (p_{11},p_{10},p_{01},p_{00})
  = (\pi\mathrm{Se},\, (1-\pi)(1-\mathrm{Sp}),\, \pi(1-\mathrm{Se}),\, (1-\pi)\mathrm{Sp}),
\end{equation*}
for $(X,Y)\in\{(1,1),(1,0),(0,1),(0,0)\}$. These cell probabilities refer to true positive, false positive, false negative, and true negative rates, respectively.

By Bayes' rule,
\begin{align*}
    \mathrm{PPV} &= \Pr(Y=1\mid X=1) = \frac{\mathrm{Se}\,\pi}{\mathrm{Se}\,\pi+(1-\mathrm{Sp})(1-\pi)}, \tag{PV1}\label{eq:ppv-bayes}\\
    \mathrm{NPV} &= \Pr(Y=0\mid X=0) = \frac{\mathrm{Sp}(1-\pi)}{(1-\mathrm{Se})\,\pi+\mathrm{Sp}(1-\pi)}. \tag{PV2}\label{eq:npv-bayes}
\end{align*}

We then define the conditional risks of clinical endpoint 
\begin{equation*}
    p_1 = \Pr(Y=1\mid X=1) = \mathrm{PPV},\qquad p_0 = \Pr(Y=1\mid X=0)=1-\mathrm{NPV}.
\end{equation*}
Thus $\psi=(p_1,p_0)^\top$ contains the risks in test-positive and test-negative strata. Unless noted otherwise, this is the meaning of $\psi$ throughout Sections~\ref{sec:variance_framework} and \ref{sec:wald}, so we write $g(\psi)$ and $g(p_1,p_0)$ interchangeably.

Write $\tau=\Pr(X=1)=p_{11}+p_{10}=\pi\Se+(1-\pi)(1-\Sp)$ for the test-positive fraction. Averaging the two stratum risks over the test result then returns the prevalence exactly,
\begin{equation}
\label{eq:marginal-identity}
    \tau p_1+(1-\tau)p_0=\Pr(Y=1)=\pi,
\end{equation}
by the law of total probability. Identity~\eqref{eq:marginal-identity} is used repeatedly below: it makes the expected event fraction equal to the design marginal, so the event-based sample size and the minimum-events floor are direct functions of $\pi$ (Section~\ref{sec:practical_considerations}).

\subsection{Effect Sizes as Risk Contrasts}

We define the estimand as
\begin{equation}
\label{eq:delta}
  \Delta = g(\psi) = f(p_1) - f(p_0),
\end{equation}
where $f(\cdot)$ sets the working scale. For non-identity links (e.g., log, logit, cloglog), we assume $p_x\in(\epsilon,1-\epsilon)$ for some small $\epsilon>0$ so derivatives are finite. When $f$ is strictly increasing, larger $p_1-p_0$ leads to larger $\Delta$.

\paragraph{Binary outcomes.}
For binary outcomes, common choices of $f(\cdot)$ yield familiar measures \cite{murphy_relationship_1983,pepe_limitations_2004}:
\[
\begin{array}{lcl}
\text{Risk difference (RD):} & f(p)=p, & \Delta_{\mathrm{RD}} = p_1 - p_0,\\[3pt]
\text{Log risk ratio (logRR):} & f(p)=\log p, & \Delta_{\mathrm{logRR}} = \log(p_1) - \log(p_0),\\[3pt]
\text{Log odds ratio (logOR):} & f(p)=\mathrm{logit}(p)=\log\!\left(\dfrac{p}{1-p}\right), &
   \Delta_{\mathrm{logOR}} = \mathrm{logit}(p_1)-\mathrm{logit}(p_0).
\end{array}
\]

\paragraph{Survival outcomes.}
For survival settings, we use a fixed follow-up horizon $t^\ast$ and define cumulative risk
\begin{equation*}
    p_x=\Pr(Y=1\mid X=x)=F_x(t^\ast)=1-S_x(t^\ast).
\end{equation*}
Under a proportional-hazards interpretation at this horizon, $S_1(t^\ast)=S_0(t^\ast)^{\mathrm{HR}}$ implies
\begin{equation*}
    \Delta_{\mathrm{logHR}} = \log[-\log(1-p_1)] - \log[-\log(1-p_0)],
\end{equation*}
which corresponds to the complementary log-log (cloglog) link $f(p)=\log[-\log(1-p)]$.

This fixed-horizon transformation preserves the risk-contrast target used for design but does not use full event-time ordering. Hence, classic continuous and discrete time-to-event analysis remains preferable for inference \cite{prentice_regression_1978,tan_analysis_2022,bottai_modeling_2021}. Nonetheless, using a time-collapsed transformation of time-to-event outcomes enables a single framework of predictive performance across outcome types within the same information-variance framework introduced later.

\subsection{Variance of the Risk-Based Effect Sizes}
\label{sec:var-risk}
Let $(n_{11},n_{10},n_{01},n_{00})$ denote the observed $2\times2$ table with total $N$. Throughout we assume population-based (cohort or cross-sectional) sampling, in which $N$ is fixed and the table is multinomial, so that $N_+\sim\mathrm{Bin}(N,\tau)$ and the stratum risks are directly estimable. In contrast, under case--control sampling, $(p_1,p_0)$ have to be reconstructed from $(\widehat{\Se},\widehat{\Sp})$ and an external $\pi$, and the two estimates are generally correlated at order $O(N^{-1})$. Define stratum sizes $N_+=n_{11}+n_{10}$ and $N_-=n_{01}+n_{00}$, and estimators $\widehat{p}_1=n_{11}/N_+$, $\widehat{p}_0=n_{01}/N_-$. Conditionally on $(N_+,N_-)$, the stratum log-likelihood is
\begin{equation*}
\ell(p_1,p_0)=
n_{11}\log p_1+(N_+-n_{11})\log(1-p_1)+
n_{01}\log p_0+(N_--n_{01})\log(1-p_0).
\end{equation*}
The conditional Fisher information is therefore diagonal. With $\tau$ as in Section~\ref{sec:from-metrics-to-effect} and using $N_+\approx\tau N$, $N_-\approx(1-\tau)N$, we obtain the information per subject. Throughout, $\mathcal{I}$ and the variances derived from it are per-subject quantities; for a sample of size $N$ the total information is $N\mathcal{I}_{\psi}$. Thus
\begin{equation}
\label{eq:Ipsi}
\mathcal{I}_{\psi}=
  \begin{pmatrix}
    \tau/[p_1(1-p_1)] & 0\\[3pt]
    0 & (1-\tau)/[p_0(1-p_0)]
  \end{pmatrix}.
\end{equation}

For $\Delta=g(\psi)$, a first-order Taylor expansion gives
\begin{equation*}
\widehat{\Delta}-\Delta
=
\nabla_\psi g(\psi)^\top(\widehat{\psi}-\psi)+O_p(N^{-1}),
\end{equation*}
with
\begin{equation*}
\nabla_\psi g(\psi)=\big(f'(p_1),-f'(p_0)\big)^\top.
\end{equation*}
Hence, by the delta method,
\begin{equation}
\label{eq:var-delta}
  \Var(\widehat{\Delta}) \approx \frac{\nabla_\psi g(\psi)^\top \mathcal{I}_{\psi}^{-1}\nabla_\psi g(\psi)}{N}.
\end{equation}
The derivative terms are
\begin{equation*}
f'(p)=
\begin{cases}
1, & \text{for RD},\\
1/p, & \text{for logRR},\\
1/[p(1-p)], & \text{for logOR},\\
1/\{(1-p)[-\log(1-p)]\}, & \text{for logHR}.
\end{cases}
\tag{D1}\label{eq:fprime}
\end{equation*}

\paragraph{Cross-stratum covariance and approximation order.}
By the law of total covariance under the population-based sampling design,
\begin{equation*}
    \mathrm{Cov}(\widehat{p}_1,\widehat{p}_0) = \mathbb{E}\!\big[\mathrm{Cov}(\widehat{p}_1,\widehat{p}_0 \mid N_+,N_-)\big] + \mathrm{Cov}\!\big(\mathbb{E}[\widehat{p}_1\mid N_+],
                   \mathbb{E}[\widehat{p}_0\mid N_-]\big).
\end{equation*}
The first term is zero because the two risks are estimated from disjoint strata conditional on $(N_+,N_-)$. The second is zero because $\mathbb{E}(\widehat{p}_1\mid N_+)=p_1$ and $\mathbb{E}(\widehat{p}_0\mid N_-)=p_0$ do not depend on the stratum sizes, provided both strata are non-empty. Random stratum sizes therefore affect the variance but not the covariance, through expansions such as
\begin{equation*}
\frac{1}{N_+}
=
\frac{1}{\tau N}
-\frac{N_+-\tau N}{\tau^2N^2}
+O_p(N^{-2}),
\end{equation*}
with an analogous expression for $1/N_-$, so that $\mathbb{E}(1/N_+)=1/(\tau N)+O(N^{-2})$. On the log scales $f(\widehat{p}_x)$ has bias of order $1/N_x$, and $\Cov(1/N_+,1/N_-)=O(N^{-3})$, so $\Cov\{f(\widehat{p}_1),f(\widehat{p}_0)\}=O(N^{-3})$. This is negligible against the $O(N^{-1})$ variance terms used for Wald-type design calculations.
This argument is also supported by finite-sample simulation in Appendix~\ref{app:fisher-info}. Across the simulated $(\Se,\Sp,\pi,N)$ grid under the population-based sampling, the closed-form $\Cov(\widehat{\mathrm{PPV}},\widehat{\mathrm{NPV}})$ is zero and the empirical values are indistinguishable from simulation noise at every $N$.

Substituting \eqref{eq:Ipsi} into \eqref{eq:var-delta} gives
\begin{equation}
\label{eq:vardelta-final}
    \Var(\widehat{\Delta}) \approx
    [f'(p_1)]^2\frac{p_1(1-p_1)}{N\tau}
    +
    [f'(p_0)]^2\frac{p_0(1-p_0)}{N(1-\tau)}.
\end{equation}

This is a first-order large-sample approximation. Near separation (e.g., $\widehat{p}_x$ close to 0 or 1), very sparse cells, or extreme prevalence, Wald variances can be anticonservative; continuity corrections and penalized estimators (e.g., Firth-type methods) are then advisable \cite{heinze_solution_2002} (see further discussion in Section~\ref{sec:practical_considerations}).

\section{Power and Sample Size for Risk-Based Effect Sizes}
\label{sec:wald}
\subsection{Hypothesis setup, Wald Statistics, Power and Sample Size}

Let $\Delta=g(p_1,p_0)$ denote the target risk contrast. We test
\begin{equation*}
H_0:\Delta=\Delta_0
\qquad\text{vs}\qquad
H_1:\Delta=\Delta_1,\ \Delta^\ast=\Delta_1-\Delta_0,
\end{equation*}
with two-sided type-I error $\alpha$ (default $\Delta_0=0$).

It is convenient to work on a per-subject scale. Write
\begin{equation}
\label{eq:v-persubject}
v(\psi)=\nabla_\psi g(\psi)^\top \mathcal{I}_{\psi}^{-1}\nabla_\psi g(\psi)
      = [f'(p_1)]^2\frac{p_1(1-p_1)}{\tau}+[f'(p_0)]^2\frac{p_0(1-p_0)}{1-\tau},
\end{equation}
which is Equation~\eqref{eq:vardelta-final} with the factor $1/N$ removed, so that $\Var(\widehat{\Delta})\approx v(\psi)/N$. We write $v$ for its value at the design configuration and keep the argument only where a different configuration is meant. Equation~\eqref{eq:v-persubject} is the single quantity that carries the design information and determines the noncentrality parameter and the sample size formula below, as well as the design efficiency of Section~\ref{sec:design_optimality}.

Inference and design in this paper use the Wald statistic standardized by its estimated standard error,
\begin{equation*}
Z_N=\frac{\widehat{\Delta}-\Delta_0}{\sqrt{\widehat{v}/N}},
\end{equation*}
where $\widehat v$ is the plug-in variance at $\widehat\psi$ (because this is the quantity which a prespecified analysis computes). Under $H_0$, $Z_N\overset{a}{\sim}\mathcal{N}(0,1)$. Under $H_1$, $\widehat v\overset{p}{\to}v$, so $Z_N$ is approximately normal with unit variance and noncentrality parameter
\begin{equation*}
\lambda_N=\frac{\sqrt{N}\,\Delta^\ast}{\sqrt{v}}.
\end{equation*}
Power at $\alpha$ is
\begin{equation*}
1-\beta \approx
\Phi\!\left(-z_{1-\alpha/2}+\lambda_N\right)+
\Phi\!\left(-z_{1-\alpha/2}-\lambda_N\right),
\end{equation*}
where $\Phi$ denotes the standard normal cumulative distribution function and the second term is negligible in practice.

Inverting the power expression above gives the sample-size formula used throughout this paper,
\begin{equation}
\label{eq:N-formula}
N = \left[ \frac{(z_{1-\alpha/2}+z_{1-\beta})\sqrt{v}}{|\Delta^*|} \right]^2,
\end{equation}
a first-order approximation consistent with standard Wald design formulas \cite{shieh_power_2005,chow_sample_2017,wang_sample_2018}. Both inputs are alternative-side quantities obtained from $(\Se,\Sp,\pi)$ by the composition of Sections~\ref{sec:from-metrics-to-effect}--\ref{sec:var-risk}, so no null configuration has to be specified, and \eqref{eq:N-formula} is the sample size used in the plug-in Wald test, in the simulations of Section~\ref{sec:empirical-validation}, and in the efficiency surfaces of Section~\ref{sec:design_optimality}.

\subsection{Practical considerations of sample size calculations for predictive test validation}
\label{sec:practical_considerations}
Equation~\eqref{eq:N-formula} is evaluated at the $(p_1,p_0,\tau)$ implied by the design inputs, giving the required sample size for a target contrast $\Delta^\ast$. This subsection recasts that size in terms of event counts and then adds the two safeguards that govern it in small studies.

\paragraph{Event-based formulation.}
Let $D=Np_{\mathrm{eff}}$ denote the effective event count with $p_{\mathrm{eff}}=\tau p_1+(1-\tau)p_0$ the expected event fraction. By identity~\eqref{eq:marginal-identity}, $p_{\mathrm{eff}}\equiv\pi$, so $D=N\pi$ at the design stage with no auxiliary quantity and Schoenfeld's event-based formula becomes a direct function of $\pi$. Equation~\eqref{eq:vardelta-final} can be written as
\begin{equation*}
\Var(\widehat{\Delta})
\approx
    \frac{p_{\mathrm{eff}}}{D}
    \left\{
    [f'(p_1)]^2\frac{p_1(1-p_1)}{\tau}
    +
    [f'(p_0)]^2\frac{p_0(1-p_0)}{1-\tau}
    \right\}.
\end{equation*}
Precision of the risk contrast is driven by the effective event count $D$ and the stratum balance through $\tau$.

For time-to-event outcomes at fixed $t^\ast$, $D$ corresponds to expected failures by $t^\ast$. The identity $D=Np_{\mathrm{eff}}$ assumes administrative censoring at $t^\ast$ with complete follow-up; under non-administrative censoring, the observed event fraction is lower, and $D$ should be inflated accordingly. Under small-to-moderate cumulative risks \emph{and} comparable stratum risks ($p_1\approx p_0\approx p_{\mathrm{eff}}$), the cloglog contrast yields the familiar proportional-hazards scaling
\begin{equation*}
    \Var(\widehat{\Delta}_{\LHR}) \propto \frac{1}{D\,\tau(1-\tau)}
\end{equation*}
(Appendix~\ref{app:schoenfeld}) \cite{schoenfeld_sample-size_1983,schoenfeld_asymptotic_1981,andersen_coxs_1982}. Both approximations are needed; the second fails precisely when the two predictive strata separate strongly, which is the regime of interest for a discriminating test. Appendix~\ref{app:schoenfeld} gives the correction factor that removes both.

\paragraph{Small-sample and rare-event correction.}
Rare outcomes, extreme prevalence, or near-perfect discrimination can make Wald designs unstable. We use two safeguards:
\begin{enumerate}
    \item \textbf{Continuity correction.}
      Replace the stratum risks by their continuity-corrected versions $\tilde{p}_1=(\tau N p_1 + 1/2)/(\tau N + 1)$ and $\tilde{p}_0=((1-\tau) N p_0 + 1/2)/((1-\tau) N + 1)$ \cite{plackett_continuity_1964}, and re-evaluate Equation~\eqref{eq:N-formula} at $(\tilde p_1,\tilde p_0)$, which corrects both $\Delta^\ast$ and $v$ through \eqref{eq:v-persubject}. The corrected risks depend on $N$, so they are evaluated at the uncorrected Wald size and the expression resolved once; this yields $N_{\mathrm{CC}}$.
    \item \textbf{Minimum information threshold.}
      Impose a minimum expected-events criterion $D_{\mathrm{eff}}=Np_{\mathrm{eff}}\ge ck$, where $k$ is the number of model parameters (typically $k=1$ here) and $c\in[5,10]$ \cite{van_smeden_no_2016}. By \eqref{eq:marginal-identity} this reduces to $N_{\mathrm{EPV}}=\lceil ck/\pi\rceil$. Near this bound, penalized methods, such as Firth's correction, should be prespecified \cite{heinze_solution_2002}.
\end{enumerate}
Here $N_{\mathrm{Wald}}$ and $N_{\mathrm{CC}}$ are evaluated from the same expression, Equation~\eqref{eq:N-formula}, at the design risks and at the continuity-corrected risks respectively; $N_{\mathrm{EPV}}$ is an independent floor on expected events rather than a variance calculation. Our default choice is $c=10$; a less conservative $c=5$ is appropriate when a companion analysis will use penalized estimation.

The final design size is then taken as
\begin{equation}
\label{eq:n_final}
    N_{\mathrm{final}} =\max\big\{N_{\mathrm{Wald}}, N_{\mathrm{CC}},N_{\mathrm{EPV}}\big\},
\end{equation}
which is our default design rule and reduces to \eqref{eq:N-formula} when both safeguards are inactive, which typically holds in confirmatory validation with abundant events and moderate risks. It remains an analytic approximation, not an exact finite-sample formula.

\paragraph{Uncertain inputs.}
The design characteristics $(\Se,\Sp,\pi)$ are rarely known exactly, but because \eqref{eq:N-formula} and \eqref{eq:n_final} are closed-form they can be evaluated directly over a grid of plausible values, for example $\Se\in[\Se_L,\Se_U]$, $\Sp\in[\Sp_L,\Sp_U]$, $\pi\in[\pi_L,\pi_U]$, giving an uncertainty envelope for required $N$ at no modeling cost. Where a joint confidence region for $(\Se,\Sp)$ is available from pilot data it can be propagated directly; otherwise marginal ranges are conservative provided grid corners are included. Section~\ref{sec:input-uncertainty} carries this out for the case study.

\subsection{Empirical validation of the variance, power, and sample-size formula}
\label{sec:empirical-validation}

\paragraph{Design.}
We evaluated finite-sample performance for binary and time-to-event estimands over a factorial grid:
\begin{equation*}
    \pi \in \{0.05, 0.10, 0.20, 0.50\},
    \qquad
    \mathrm{Se},\mathrm{Sp} \in \{0.60, 0.80, 0.95\},
    \qquad
    \text{target power} \in \{0.80, 0.90\},
\end{equation*}
with independent administrative censoring at time $t^*$ and optional additional censoring at rate $c\in\{0,0.20\}$ for survival settings. For each configuration, $(p_1,p_0)$ and $\tau$ were computed from $(\pi,\Se,\Sp)$ (Section~\ref{sec:from-metrics-to-effect}). Total sample size $N$ was set by Equation~\eqref{eq:N-formula}, then updated using Equation~\eqref{eq:n_final} when continuity-correction or EPV constraints were active.

\paragraph{Data generating process.}
Since our framework aims to provide a practical design workflow for both binary and survival endpoints, all outcome data were simulated from a Weibull$(\kappa)$ distribution and calibrated to the design risks at $t^*$. Given $(p_0,p_1)$, baseline hazards for group $x\in\{0,1\}$ satisfy
\begin{equation*}
    \lambda_x = -\log(1-p_x)\big/t^{*\kappa},
\end{equation*}
so that $\Pr(T\le t^*\mid X=x)=p_x$. By identity~\eqref{eq:marginal-identity} the $\tau$-weighted mixture of the two stratum risks is $\pi$, so the design marginal is reproduced by construction.

Each subject was assigned a latent status $D\sim\mathrm{Bernoulli}(\pi)$ and then a test result from the operating characteristics, $X\mid D=1\sim\mathrm{Bernoulli}(\Se)$ and $X\mid D=0\sim\mathrm{Bernoulli}(1-\Sp)$, so that $\Pr(X=1)=\tau$ marginally; event times were drawn within the resulting test strata. Optional independent censoring was added via $C\sim\mathrm{Exp}(\lambda)$ with $\lambda=-\log(1-\Pr(C\le t^*))/t^*$. Observed times were $Y=\min(T,C,t^*)$, event indicators $d=\mathds{1}\{T\le C,\,T\le t^*\}$, and observed event count $D=\sum d$. Each parameter combination used 2000 replicates.

\paragraph{Analysis and metrics.}
Binary estimands were analyzed with plug-in Wald estimators, adding $1/4$ to each cell of the observed $2\times 2$ table when its smallest count was at most one. (This analysis-stage correction is deliberately lighter than the $1/2$ used for design in Section~\ref{sec:practical_considerations}, which is applied to expected rather than observed counts.) Survival estimands used (a) empirical cloglog contrast at $t^*$, (b) Cox partial-likelihood estimation, and (c) Cox estimation with Firth correction. We report relative bias (bias divided by the true effect), SE calibration (the empirical standard deviation of the estimate divided by the mean analytic standard error, so that one denotes exact calibration), 95\% Wald confidence-interval coverage, and achieved power.

\paragraph{Results.}
The simulation grid spanning $(\pi,\Se,\Sp)$ generated a wide range of effect sizes: RD values from 0.04 to 0.90, logRR from 0.41 to 5.20, logOR from 0.81 to 5.89, and logHR from 0.58 to 5.52. After applying the extreme-case corrections described earlier, the corresponding sample size requirements ranged from 20 to 1508, depending on the estimand and its magnitude.

\begin{figure}
    \centering
    \includegraphics[width=1\linewidth]{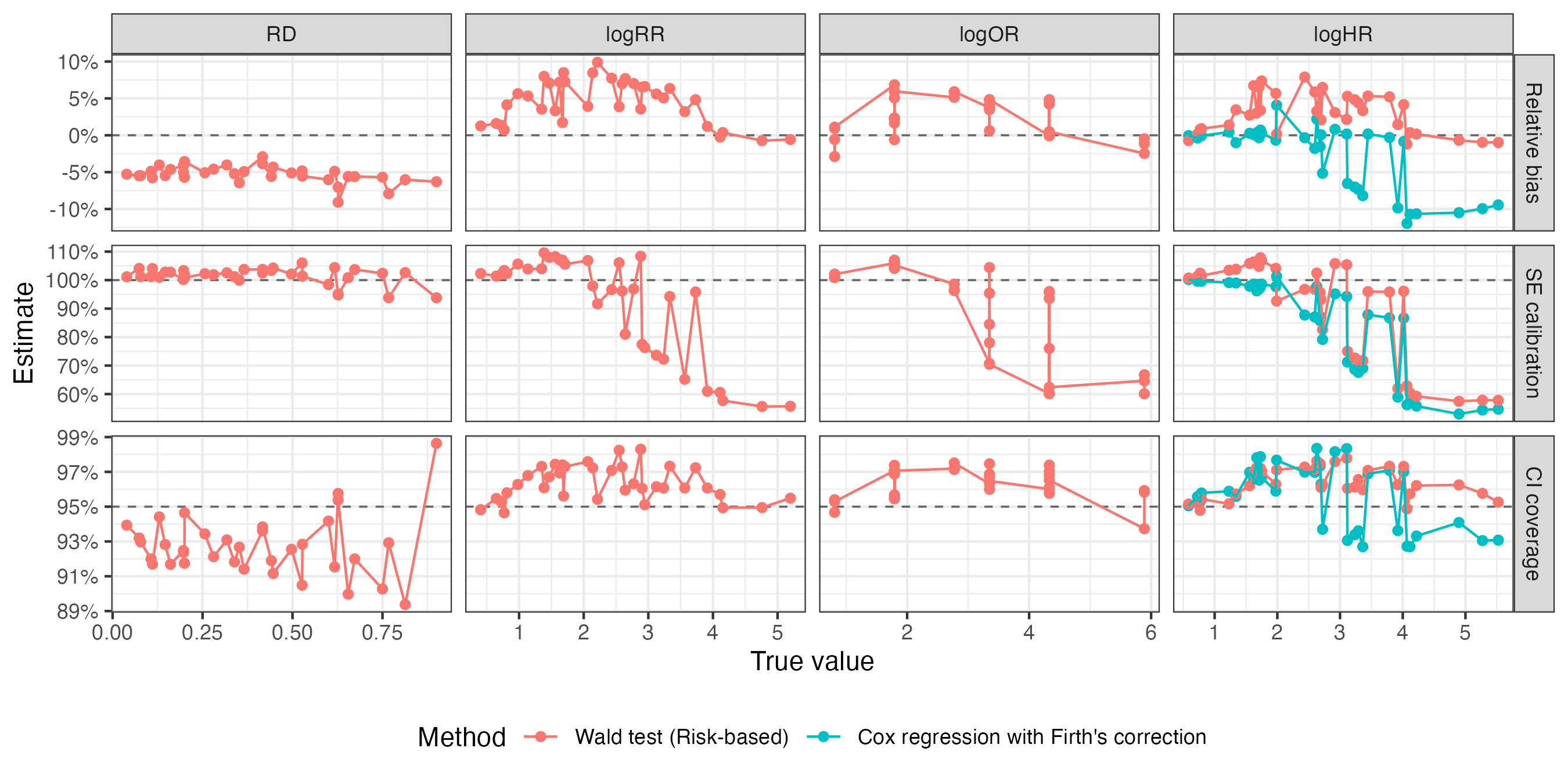}
    \caption{Relative bias, SE calibration, and 95\% CI coverage for binary and survival estimands. SE calibration is the empirical standard deviation of the estimate divided by the mean analytic standard error, so the dashed reference line at $100\%$ marks exact agreement; values below it indicate that Equation~\eqref{eq:vardelta-final} overstates sampling variability. Dashed lines in the other rows mark zero bias and nominal $95\%$ coverage. Each point pools the two target power levels and the two censoring fractions for that configuration. For logOR, the same true value can arise from multiple $(\Se,\Sp,\pi)$ combinations, so multiple empirical estimates may align with one true logOR value.}

    \noindent\textit{Alt text: Three rows of scatter panels (relative bias, SE calibration ratio, 95\% CI coverage) plotted against true effect size, one column per estimand (RD, logRR, logOR, logHR); points track their reference lines closely at small-to-moderate effect sizes and diverge at large effect sizes.}
    \label{fig:bias_plot}
\end{figure}

Figure~\ref{fig:bias_plot} shows relative bias, SE calibration, and 95\% CI coverage as functions of the true effect size. For binary estimands (RD, logRR, logOR), the empirical relative bias of the marginal risk--based estimators (Equation~\eqref{eq:delta}, with continuity correction when needed) was generally small across the design grid, within $\pm 10\%$ throughout. The analytic standard error of Equation~\eqref{eq:vardelta-final} tracked the sampling standard deviation closely over the small-to-moderate effect range, with SE calibration ratios near one (median $0.96$--$1.02$ by estimand). Wald-type 95\% confidence intervals achieved close-to-nominal coverage across most configurations, running slightly conservative on the log scales ($94$--$98\%$) and mostly slightly anticonservative for RD ($89$--$96\%$), where the small negative bias in $\widehat{\Delta}_{\RD}$ is largest relative to its standard error.

At larger effects the analytic standard error becomes conservative rather than optimistic: the calibration ratio falls to roughly $0.6$ on the log scales once the true contrast exceeds about $2.5$, meaning Equation~\eqref{eq:vardelta-final} overstates sampling variability by up to about $1.8$-fold there. This is the mechanism behind the mild overpowering reported below, and it is the direction a design calculation should prefer.

For survival estimands, both the empirical cloglog estimator at $t^*$ and the Cox model with Firth correction remained close to unbiased across censoring fractions, the latter drifting to about $-10\%$ relative bias with coverage near $93\%$ at the largest effects, consistent with the shrinkage that penalisation induces. The standard Cox partial-likelihood estimator without correction behaved very differently. Relative bias rose from under $2\%$ at $\LHR\approx1.2$ to about $+100\%$ at $\LHR\approx2.4$ and above $+200\%$ beyond $\LHR\approx3$. In these settings, risk sets became highly imbalanced and near-separation arose, which inflated the model-based standard error far more than the sampling variability, and its calibration ratio collapses towards zero (Figure~\ref{fig:app_bias_plot}). Coverage therefore remained nominal or above, because the intervals were far too wide rather than too narrow; the cost appears as a loss of power rather than as undercoverage.

\begin{figure}
    \centering
    \includegraphics[width=1\linewidth]{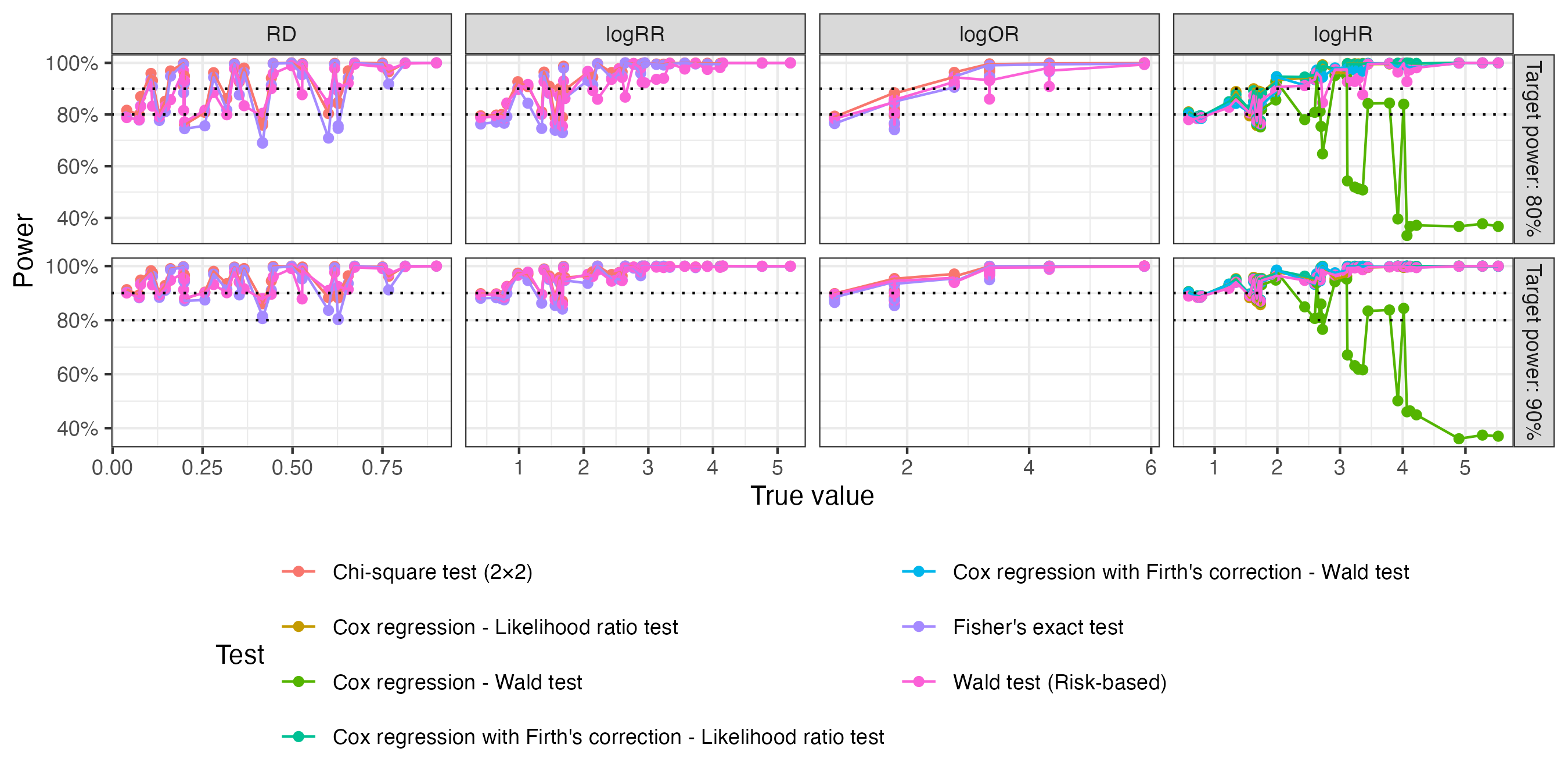}
    \caption{Empirical power for binary and survival estimands. Power is shown as a function of true effect size at target levels of 80\% (top row) and 90\% (bottom row). Most tests achieve nominal power across estimands, except for the uncorrected Cox regression Wald test.}

    \noindent\textit{Alt text: Empirical power curves versus true effect size for RD, logRR, logOR, and logHR at 80\% and 90\% target power; most curves track the nominal power target while the uncorrected Cox Wald test drops sharply at large logHR.}
    \label{fig:power_plot}
\end{figure}

Power results showed a similar pattern (Figure~\ref{fig:power_plot}). Wald tests based on empirical risks (with continuity correction when required) achieved their nominal power levels (80\% or 90\%) when $\LHR\lesssim 1.5$. All evaluated tests tended to be slightly overpowered at larger effects. The analytic standard error is conservative there, so the realised signal-to-noise ratio exceeds the design target. Likelihood ratio and Firth-corrected Cox tests showed comparable behavior. The only systematic deviation was the Wald test from the uncorrected Cox regression, which became markedly underpowered when $\LHR\gtrsim 3$.

Across this grid, the bias and calibration summaries show that the current Wald-type variance and sample-size expressions are well calibrated in small-to-moderate effect regimes, including rare-disease prevalence ($\pi=0.05$), and err towards conservatism rather than optimism outside them. Deviations occurred primarily in high-effect and sparse-data settings, where separation and extreme imbalance undermine standard Wald approximations. These findings support the use of continuity and EPV safeguards in the conservative setting of pilot study design \cite{van_smeden_no_2016}. The cross-term between $\widehat{p}_1$ and $\widehat{p}_0$ remained negligible for design-level variance calculations (Appendix~\ref{app:fisher-info}).

\section{Design Efficiency for Resource-Constrained Pilot and Feasibility Studies}
\label{sec:design_optimality}
Pilot and feasibility studies ($N\approx 40$--$150$) must both estimate Se/Sp and detect an early effect-size signal, and their operating point can still be tuned before the classifier is locked (via threshold along the ROC and via eligibility criteria that shift $\pi$). These studies rarely have infrastructure for large simulation campaigns, so analytic design surfaces are practical.

We therefore adopt optimal experimental design theory and use C-optimal and D-optimal surfaces below as decision tools for pilot protocol planning. The default sample-size recommendation in this setting is Equation~\eqref{eq:n_final}, not the uncorrected Wald size. The max rule is more stable in small studies and collapses to the Wald formula when continuity and EPV safeguards are inactive.

\subsection{Feasible Design Region and Limits}
\label{subsec:bounds_effectsizes}
Before using optimization surfaces, we define the admissible design region $\Omega$ for $(\Se,\Sp,\pi)$ and the implied risk contrasts (Equation~\eqref{eq:delta}).

Assume
\begin{equation*}
    \Omega =
    \left\{
        (\Se,\Sp,\pi):
        0 < \pi < 1,\;
        0 < \Se < 1,\;
        0 < \Sp < 1,\;
        \Se+\Sp>1
    \right\},
\end{equation*}
so the test performs better than chance (positive Youden index for a reasonable test).
Within $\Omega$, $\Delta$ is constrained by the mapping
\begin{equation*}
    \Delta
    = f\!\big(p_1(\Se,\Sp,\pi)\big)-f\!\big(p_0(\Se,\Sp,\pi)\big).
\end{equation*}

\paragraph{Extremal values for the risk difference.}
For RD, $f(p)=p$ and $\Delta_{\RD}=p_1-p_0$. Pilot data typically establish that a test discriminates better than chance by some margin. Let $J_0\in(0,1]$ denote a minimum Youden index (equivalently, $\Se+\Sp\ge 1+J_0$), and define the constraint set $\Omega_{J_0}=\{(\Se,\Sp): \Se+\Sp\ge 1+J_0\}$. At fixed prevalence $\pi$,
\begin{equation}
    \label{eq:delta_maxmin}
  \Delta_{\max}(\pi,J_0) = \max_{(\Se,\Sp)\in\Omega_{J_0}}\big[p_1(\Se,\Sp,\pi)-p_0(\Se,\Sp,\pi)\big],
  \quad
  \Delta_{\min}(\pi,J_0) = \min_{(\Se,\Sp)\in\Omega_{J_0}}\big[p_1(\Se,\Sp,\pi)-p_0(\Se,\Sp,\pi)\big].
\end{equation}
Because $\Delta_{\RD}$ is increasing in both $\Se$ and $\Sp$, the supremum is attained in the limit $(\Se,\Sp)\to(1,1)$, and the infimum is attained on the constraint boundary $\Se+\Sp=1+J_0$. Extremizing along this boundary yields the closed-form bounds
\begin{equation}
    \label{eq:delta_bounds}
        4\pi(1-\pi)J_0 \;\le\; \Delta_{\RD} \;\le\; 1,
\end{equation}
with the lower bound attained at $\Se^\ast=1/2+(1-\pi)J_0$, $\Sp^\ast=1/2+\pi J_0$ (balanced test-positive rate $\tau=1/2$), provided $\max(\pi,1-\pi)\,J_0 \le 1/2$. Otherwise the minimum is attained at a boundary endpoint of the constraint segment. The lower bound depends jointly on prevalence and on the minimum discrimination the test is known to achieve, defining an admissible target region for feasibility screening.

The same mapping applies to logRR, logOR, and logHR, but RD is useful for feasibility checks because it is bounded on the probability scale. Ratio-based contrasts may diverge near boundaries and are less informative for defining admissible target regions.

\paragraph{Variance bounds for risk difference.}
From Equation~\eqref{eq:vardelta-final}, with $p_1,p_0\in[0,1]$,
\begin{equation}
        0\le \mathrm{Var}(\widehat{\Delta}_{\RD})
        \le \frac{1}{4N}\!\left(\frac{1}{\tau}+\frac{1}{1-\tau}\right) 
        = \frac{1}{4N}\!\left[\frac{1}{\pi\mathrm{Se}+(1-\pi)(1-\mathrm{Sp})}
            +\frac{1}{\pi(1-\mathrm{Se})+(1-\pi)\mathrm{Sp}}\right].
    \label{eq:var-envelope}
\end{equation}

The upper envelope is a conservative worst case, tight only at $p_1=p_0=0.5$ (a null-effect configuration) and loose elsewhere, and it is minimized in $\tau$ at $\tau=1/2$. 

\subsection{Design Efficiency using C-optimality as a Pilot Decision Tool}
\label{subsec:copt}
When a single estimand is primary (e.g., RD or logHR), C-optimal design in the optimal experimental design theory targets its precision \cite{pukelsheim_optimal_2006}. From Equation~\eqref{eq:var-delta}, this precision is governed by $\mathcal{I}_{\psi}$. Using the per-subject variance \eqref{eq:v-persubject}, we define the effect-size efficiency as
\begin{equation}
\Eff = \frac{\Delta^2}{v},
\label{eq:E1}
\end{equation}
the squared contrast per unit of variance. Larger $\Eff$ implies a greater signal-to-noise ratio and thus a smaller required sample size for a fixed target effect; because $v$ is per subject, $\Eff$ is normalized for sample size and independent of $N$. Comparing \eqref{eq:E1} with \eqref{eq:N-formula} makes the relationship exact,
\begin{equation}
N=\frac{\left(z_{1-\alpha/2}+z_{1-\beta}\right)^{2}}{\Eff},
\label{eq:N-from-Eff}
\end{equation}
so the efficiency surfaces below and the required-size surfaces are the same object up to a constant that depends only on $\alpha$ and the target power. This is the sense in which test performance metrics link directly to Fisher information and optimal design principles through effect sizes.

To hold discrimination (AUC) fixed while varying threshold and prevalence (i.e., the two quantities under investigator control at design phase) we visualize the design surface under a binormal model with classifier scores $\mathcal N(0,1)$ and $\mathcal N(d,1)$ in non-event and event groups. The separation parameter is $d=\sqrt{2}\,\Phi^{-1}(\mathrm{AUC})$, where AUC is the area under the receiver operating characteristic (ROC) curve. At threshold $c$,
\begin{equation*}
    \mathrm{Sp}=\Phi(c), \qquad \mathrm{Se}=1-\Phi(c-d),
\end{equation*}
which traces the ROC trade-off.

\begin{figure}[ht]
\centering
\includegraphics[width=0.8\linewidth]{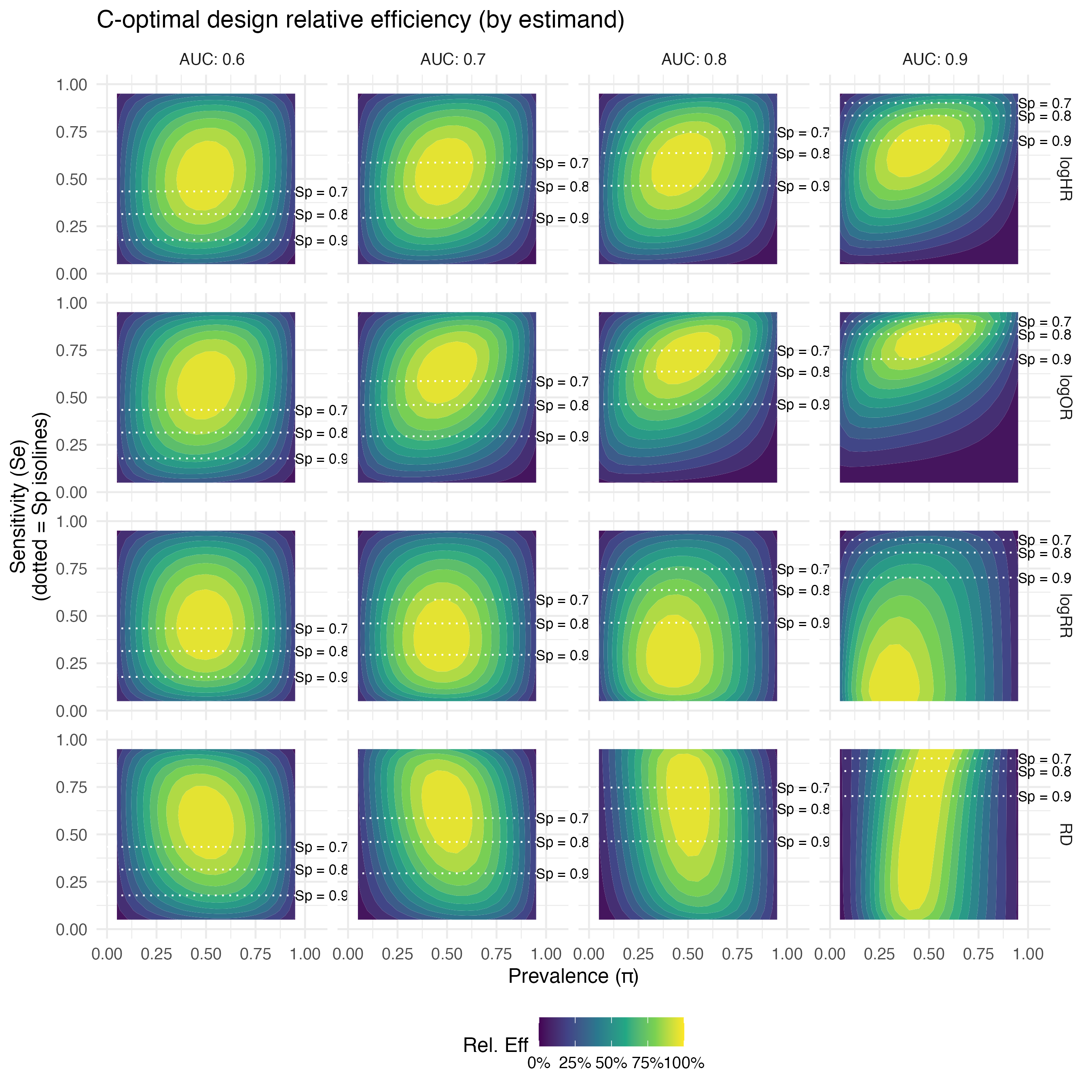}
\caption{C-optimal per-subject efficiency surface. Efficiency is shown over prevalence and sensitivity, with each column holding discrimination fixed at the stated AUC and each row giving one estimand; specificity is then determined by the ROC, and dotted lines mark specificity isolines. Values are relative to the $99.9$th percentile of efficiency within each panel, so colours are comparable within but not across panels. Interior operating points typically dominate extremes because both predictive strata contribute information.}

\noindent\textit{Alt text: Heatmap panels of C-optimal per-subject efficiency over prevalence and sensitivity, one column per AUC level and one row per estimand, with specificity isolines overlaid; efficiency is highest at interior operating points and lowest near the extremes.}
\label{fig:copt-surface_eff}
\end{figure}

\begin{figure}[ht]
\centering
\includegraphics[width=0.8\linewidth]{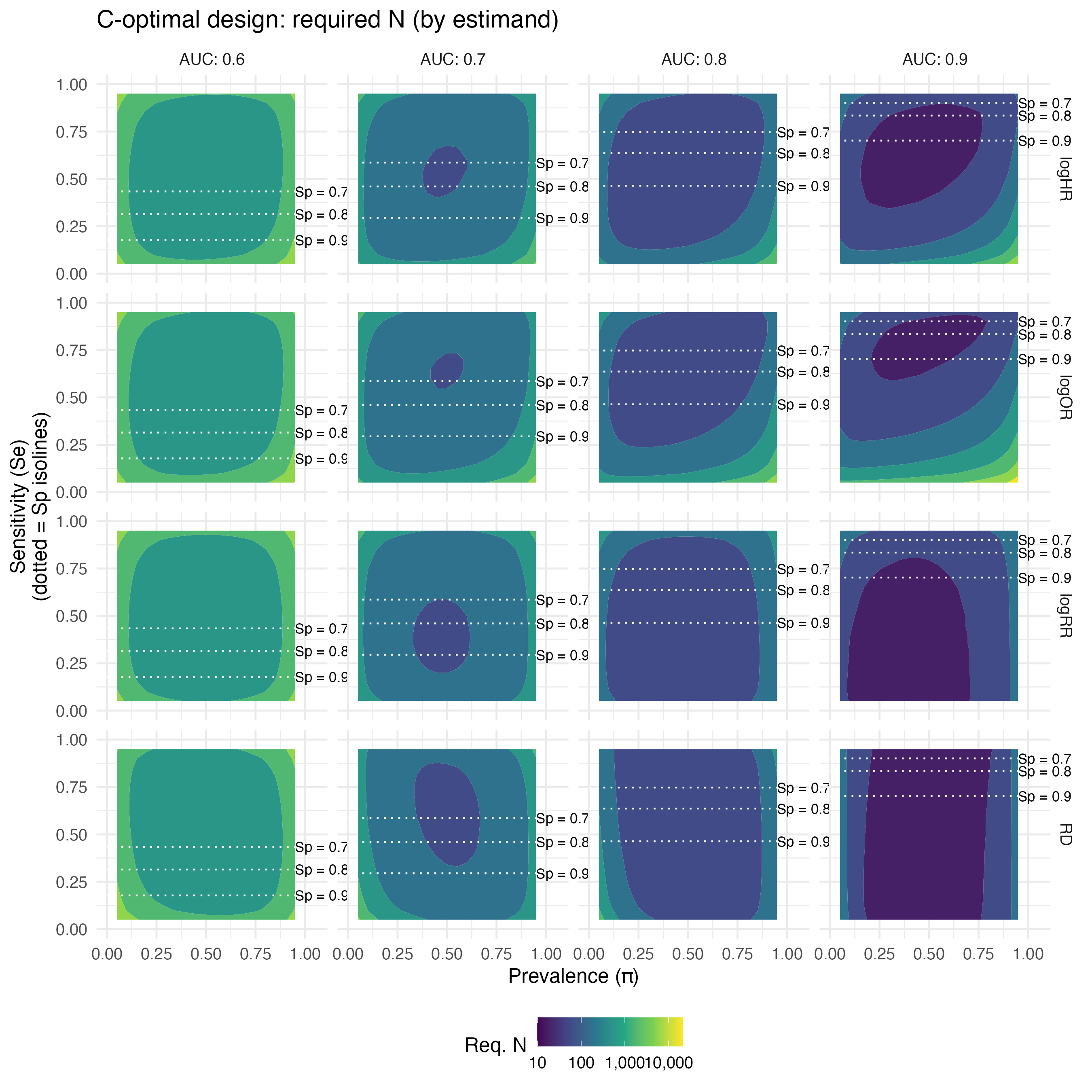}
\caption{C-optimal required sample size surface, over prevalence and sensitivity with discrimination fixed within each column. Contours of $\log_{10}(N)$ mirror the inverse of per-subject efficiency, with larger sample size required near prevalence or sensitivity extremes; the legend is labelled on the $N$ scale.}

\noindent\textit{Alt text: Contour panels of required sample size $N$ (log scale) over prevalence and sensitivity, one column per AUC level; contours show $N$ increasing sharply near prevalence or sensitivity extremes.}
\label{fig:copt-surface_N}
\end{figure}

Given a plausible discrimination level, Figures~\ref{fig:copt-surface_eff} and \ref{fig:copt-surface_N} show where small shifts in prevalence targeting or threshold choice can change required sample size by large multiples. In the $\mathrm{AUC}=0.7$, RD panel, for instance, holding prevalence at $30\%$ and moving only the threshold along the ROC takes the required size from about $110$ to about $430$.

As $\tau$ approaches $0$ or $1$, the variance Equation~\eqref{eq:vardelta-final} grows without bound, because one predictive stratum then holds almost no subjects. Prevalence enters precision through the same term. Extreme thresholds are therefore inefficient in most configurations, even when discrimination is strong. One exception matters in practice. Write $h(p)=[f'(p)]^2p(1-p)$, so that the per-subject variance \eqref{eq:v-persubject} is $v=h(p_1)/\tau+h(p_0)/(1-\tau)$. On the log scales $h$ diverges as $p\to 0$. A stratum risk close to zero can then contribute more to $v$ than the imbalance in $\tau$ does, so lifting that risk off the boundary improves efficiency even at the cost of $\tau(1-\tau)$. The CACS case study (Section~\ref{sec:casestudy}) is such a setting. No comparable exception arises for RD, where $h(p)=p(1-p)$ instead vanishes at the boundary. In either case the operating point is a clinical decision as much as a statistical one.

\subsection{D-optimal Design as a Complementary Precision Diagnostic}
\label{subsec:dopt}

While C-optimality targets one estimand, D-optimality targets joint predictive-value precision in the optimal experimental design theory framework. Let $\bphi=(\mathrm{PPV},\mathrm{NPV})^\top$. The criterion is
\begin{equation*}
    \max_{\mathcal{D}} \; |\mathcal{I}_{\bphi}|.
\end{equation*}

Because $|\mathcal{I}^{-1}_{\bphi}|^{1/2}$ is proportional to confidence-ellipsoid volume, D-optimality minimizes joint uncertainty. With negligible covariance (Appendix~\ref{app:fisher-info}),
\begin{equation*}
    |\mathcal{I}_{\bphi}| \propto \frac{1}{\mathrm{Var}(\widehat{\mathrm{PPV}})\mathrm{Var}(\widehat{\mathrm{NPV}})}.
\end{equation*}

D-optimality can favor extreme operating points, an artifact of Bernoulli variance 
vanishing near the boundaries, which matches PPV/NPV precision formulas in 
Steinberg's work (designs similarly favour high-specificity or high-sensitivity 
operating points) \cite{steinberg_sample_2009}. This boundary concentration limits 
its standalone usefulness for effect-size studies (Appendix~\ref{app:d-opt}), and 
we therefore use D-optimality as a complementary diagnostic for predictive-value 
precision, not as the primary design criterion.

\section{Case Study: Clinical validation of coronary artery calcium scoring for cardiovascular disease in type 2 diabetes}
\label{sec:casestudy}
\subsection{Clinical setting and available data}
This case study is derived from published data from a long-term cohort study of the coronary artery calcium score (CACS) in asymptomatic patients with type 2 diabetes mellitus (T2DM) \cite{koo_coronary_2023}. Patients underwent CACS evaluation at baseline (Agatston scoring by multidetector computed tomography) and were followed for incident cardiovascular disease (CVD), defined as coronary, cerebrovascular, or peripheral arterial disease, over a study period of up to 12 years (median follow-up, 10.1 years). The reported receiver operating characteristic (ROC) analysis identified a CACS threshold of 10 Agatston units as a clinically useful cutoff for stratifying patients by CVD risk.

Treating ``test positive'' as $\mathrm{CACS}\ge 10$ and the target $Y$ as incident CVD, we summarize the reported operating characteristics at this cutoff:
\begin{equation*}
\Se=0.82,\quad \Sp=0.68,\quad \pi=0.072,\quad t^\ast=12\ \text{years},\quad N_{\text{total}}=981,
\end{equation*}
where $\pi=71/981$ is the observed cumulative CVD incidence over follow-up. The publication reports $\Se=81.7\%$ (95\% CI 70.7--89.9) and $\Sp=67.9\%$ (95\% CI 64.8--70.9) at the CACS$=$10 cutoff, with 71 of 981 patients developing CVD during follow-up; we round $(\Se,\Sp)$ to two decimals for the design calculation. Because median follow-up (10.1 years) is shorter than the horizon $t^\ast=12$, the crude proportion $71/981$ understates $\Pr(Y=1\mid t^\ast)$; consistent with the caveat in Section~\ref{sec:practical_considerations}, the design $N$ derived from this crude $\pi$ is upward-biased for the intended horizon. The magnitude is not trivial: at fixed $(\Se,\Sp)=(0.80,0.70)$, Table~\ref{tab:cacs-sensitivity} shows $N$ dropping from 152 to 110 as $\pi$ moves from 0.07 to 0.10, so a Kaplan--Meier--based horizon-risk correction could reduce the required $N$ by tens of percent rather than a small fraction.

Mapping these inputs gives
\begin{equation*}
\mathrm{PPV}
=\frac{0.82\times0.072}{0.82\times0.072+(1-0.68)\times(1-0.072)}
\approx0.17,\qquad
\mathrm{NPV}
=\frac{0.68\times(1-0.072)}{(1-0.82)\times0.072+0.68\times(1-0.072)}
\approx0.98.
\end{equation*}
Therefore,
\begin{equation*}
p_1=\mathrm{PPV}\approx0.17,\qquad p_0=1-\mathrm{NPV}\approx0.02.
\end{equation*}
The implied test-positive fraction is $\tau=\pi\Se+(1-\pi)(1-\Sp)\approx0.357$, consistent with the reported proportion of patients with $\mathrm{CACS}\ge10$ (35.7\%).

\subsection{Mapping to the unified Wald framework}
For the survival contrast, we compute
\begin{equation*}
\Delta_{\LHR}
=\log[-\log(1-p_1)]-\log[-\log(1-p_0)],
\end{equation*}
then apply Equation~\eqref{eq:var-delta} for the Wald standard error.

Table~\ref{tab:cacs-compare} shows the two estimates side by side. The converted fixed-horizon log hazard ratio ($\Delta_{\LHR}\approx2.19$, $\mathrm{HR}\approx8.9$) sits close to the adjusted Cox estimate reported for $\mathrm{CACS}\ge10$ versus $<10$ ($\LHR\approx2.13$, $\mathrm{HR}=8.41$), and its standard error is slightly smaller (0.31 against 0.34). The agreement should not be over-read. Collapsing follow-up at $t^\ast$ leaves the Wald variance dependent only on binary status at that horizon, so it uses neither the event-time ordering nor the censoring pattern, and the converted contrast is marginal in $(X,Y)$ whereas the reported one is confounder-adjusted. The converted standard error also treats $(\Se,\Sp)$ as fixed, whereas the reported sensitivity carries a 95\% interval of 70.7 to 89.9\%. It is therefore a design-stage approximation rather than an inferential quantity, and an unadjusted Cox model fitted to the source cohort would be the closest directly comparable benchmark.

\begin{table}[ht]
\centering
\caption{Model-based and converted log HR for the CACS/CVD data. The reported estimate is the fully adjusted Cox model for CACS$\ge$10 versus $<$10. The converted estimate is the marginal fixed-horizon contrast implied by $(\Se,\Sp,\pi)$.}
\label{tab:cacs-compare}
\begin{tabular}{lcccc}
\toprule
Source & logHR & SE & 95\% CI (logHR) & HR (95\% CI) \\
\midrule
Reported (adjusted Cox) & 2.13 & 0.34 & [1.46, 2.80] & 8.41 [4.30, 16.46] \\
Converted (fixed-horizon) & 2.19 & 0.31 & [1.58, 2.80] & 8.92 [4.86, 16.37] \\
\bottomrule
\end{tabular}
\end{table}

\subsection{Sample-size and efficiency implications}

Using the observed operating point $(\Se,\Sp,\pi)$, Equation~\eqref{eq:N-formula} gives $N_{\mathrm{Wald}}=155$ at 80\% power and $207$ at 90\%. Applying the safeguards on the same expression (Section~\ref{sec:practical_considerations}) at 80\% power, the continuity correction gives $N_{\mathrm{CC}}=153$ and the events-per-parameter floor at $c=10$ gives $N_{\mathrm{EPV}}=\lceil c/\pi\rceil=139$, so the final size is $N_{\mathrm{final}}=\max(155,\,153,\,139)=155$ and the Wald term itself is binding. Both safeguards are nevertheless close to active, which is the characteristic signature of the rare-outcome regime: $\pi=0.072$ implies only $D=N\pi\approx11$ expected events at the design size, barely above the events-per-parameter threshold, so a modestly stronger operating point moves the design onto the EPV floor rather than off it (Table~\ref{tab:cacs-sensitivity}).

Although the hazard ratio is large ($\widehat{\Delta}_{\LHR}\approx2.19$, $\mathrm{HR}\approx8.9$), the low CVD incidence ($\pi\approx7\%$) keeps the effective event count modest ($D=N\pi\approx11$ at 80\% power). Rarity of the outcome, not weak discrimination, drives the requirement upward. With more common outcomes, both strata carry many events and required $N$ drops (at the same $(\Se,\Sp)$, $N_{\mathrm{Wald}}=64$ at $\pi=0.20$ and $39$ at $\pi=0.50$). The actual study ($N=981$, 71 events) was therefore sufficiently powered, consistent with its reported confidence intervals excluding the null ($\mathrm{HR}=8.41$, 95\% CI [4.30, 16.46]). The operating point sits above the $|\Delta_{\LHR}|\lesssim1.5$ range in which the simulations of Section~\ref{sec:empirical-validation} confirmed nominal calibration, and at $D\approx 11$ it sits just above the events-per-parameter floor. Both suggest the safeguarded rule \eqref{eq:n_final} is expected to be conservative in this regime, and any residual bias is toward over-powering.

The inputs differ from those of the classic Schoenfeld formula \cite{schoenfeld_sample-size_1983}, which starts from an anticipated hazard ratio, an allocation ratio, and an event count instead of deriving all three from $(\Se,\Sp,\pi)$. At the CACS point a naive Schoenfeld calculation returns $N\approx 100$ at 80\% power, against $N=155$ from Equation~\eqref{eq:N-formula}. Such difference comes from the correction factor $K$ of Appendix~\ref{app:schoenfeld} evaluated here, because the Schoenfeld per-subject variance $1/[\pi\tau(1-\tau)]=60.6$ is what Equation~\eqref{eq:vardelta-final} returns, when both stratum risks are set to their common marginal value $\pi$, against $v=94.1$ at the observed point.

\begin{figure}[H]
\centering
\includegraphics[width=0.7\linewidth]{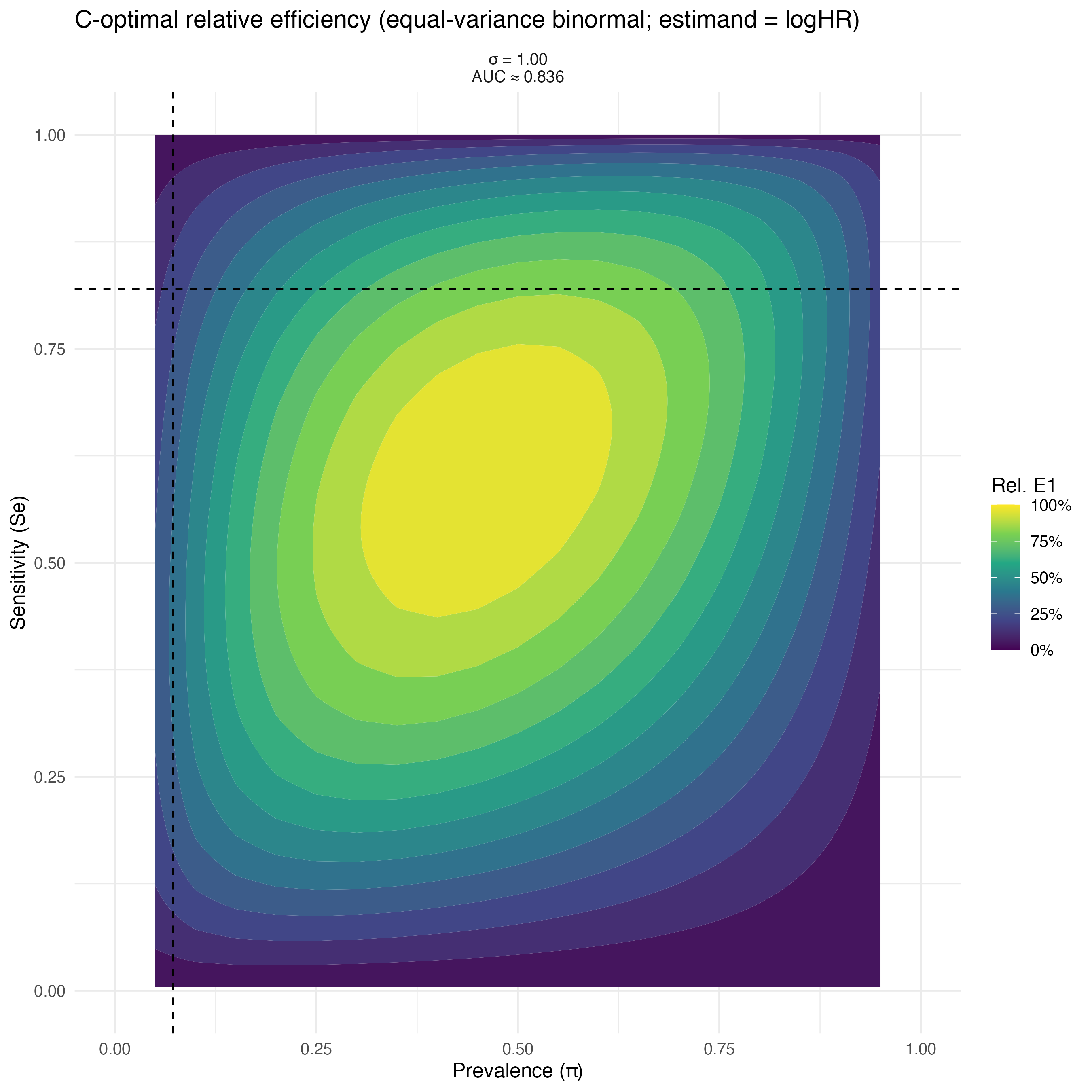}
\caption{Per-subject efficiency for $\Delta_{\LHR}$, as a percentage of the maximum attained on the surface, over prevalence $\pi$ and sensitivity under an equal-variance binormal ROC assumption anchored on the observed $(\mathrm{Se},\mathrm{Sp})=(0.82,0.68)$, giving $d\approx1.38$ and $\mathrm{AUC}\approx0.84$. Dashed lines indicate the published $\mathrm{CACS}\ge10$ operating point, which lies on the assumed ROC by construction.}
\noindent\textit{Alt text: Heatmap of per-subject efficiency for the logHR contrast over prevalence and sensitivity under a binormal ROC anchored at the published CACS operating point; the published point sits well below the efficiency maximum, which occurs at much higher prevalence.}
\label{fig:cacs-surface}
\end{figure}

To examine efficiency near the published operating point, we constructed an equal-variance binormal ROC surface anchored on $(\Se,\Sp)=(0.82,0.68)$, which implies $d=\Phi^{-1}(\Sp)-\Phi^{-1}(1-\Se)\approx1.38$ and $\mathrm{AUC}=\Phi(d/\sqrt{2})\approx0.84$. We do not anchor the surface on the AUC reported by \cite{koo_coronary_2023} ($\mathrm{AUC}=0.748$), because for a dichotomized test the reported AUC coincides with $(\Se+\Sp)/2$ and reflects the discrimination of the binary CACS$\ge$10 indicator rather than the continuous Agatston score; feeding that value into a continuous-score binormal ROC would leave the observed operating point off the assumed curve, since its Youden index ($0.50$) exceeds the maximum Youden ($0.36$) attainable on a $d\approx0.95$ curve. Varying prevalence and threshold over this curve, Figure~\ref{fig:cacs-surface} puts the published point at about 20\% of the maximum per-subject efficiency, an envelope of roughly $4.9\times$ fewer patients for the same precision. Almost all of that envelope lies along the prevalence axis. The optimum sits at $\pi\approx0.45$, $\Se\approx0.60$, $\Sp\approx0.87$, and a cohort with nearly 50\% CVD incidence is not obtainable in asymptomatic T2DM primary prevention.

The threshold is the lever an investigator can actually move. At fixed $\pi=0.072$ a threshold-only sweep peaks near $c\approx1.6$, where $(\Se,\Sp)\approx(0.42,0.94)$ and $N_{\mathrm{Wald}}$ falls from 155 to 87. A sensitivity of $0.42$ is difficult to defend for a primary-prevention risk stratifier. The gain is also smaller than it appears, because the events-per-parameter floor binds at that operating point and $N_{\mathrm{final}}$ falls only from 155 to 139. 

The direction of the efficient move is worth calling out because it runs opposite to the general claim in Section~\ref{subsec:copt}. With rare outcomes, the rare-stratum variance term dominates the balance term, and the efficient move is to raise the threshold rather than to seek an interior operating point. In practice, the binding constraints are the clinical acceptability of a lower-sensitivity cutoff and the representativeness of an enriched cohort rather than statistical efficiency alone.

\subsection{Sensitivity of design requirements to input uncertainty}
\label{sec:input-uncertainty}
The operating characteristics used in Section~\ref{sec:casestudy} were taken from a single cohort study. To assess sensitivity of the design to plausible variation in these inputs, we evaluated the sample-size formula \eqref{eq:N-formula} over a grid spanning weaker-to-stronger discrimination and lower-to-higher CVD incidence:
\begin{equation*}
\Se \in \{0.60, 0.70, 0.80, 0.85, 0.90\}, \quad
\Sp \in \{0.60, 0.70, 0.80, 0.85, 0.90\}, \quad
\pi \in \{0.05, 0.07, 0.10\}.
\end{equation*}
For each combination, the implied $(p_1,p_0)$, effect size $\Delta_{\LHR}$, and required $N$ for 80\% power (two-sided $\alpha=0.05$) were computed using the closed-form expressions. The observed operating point $(\Se,\Sp,\pi)=(0.82,0.68,0.072)$ is included as an additional row.

Table~\ref{tab:cacs-sensitivity} reports the uncorrected $N_{\eqref{eq:N-formula}}$ and safeguarded $N_{\mathrm{final}}$ for selected combinations. At fixed $(\Se,\Sp)=(0.80,0.70)$, reducing $\pi$ from 0.07 to 0.05 raises the uncorrected $N$ by about 37\% (152 to 208). On the other hand, a near-chance test ($\Se=\Sp=0.60$) requires several hundred to over a thousand subjects, while $\Se=\Sp=0.90$ falls below 80 before any safeguard is applied. The events-per-parameter floor then compresses the low end sharply at low prevalence (i.e., $N_{\mathrm{final}}=143$) because the event count rather than the risk separation becomes binding. The safeguarded $N_{\mathrm{final}}$ still spans roughly an order of magnitude across the grid (about 110 to over 1000), but strong discrimination no longer buys a proportionate reduction because of the EPV rule. Lowering $c$ to 5 when penalized estimation is planned removes the floor for every row in this table, at which point the Wald and continuity-corrected terms govern throughout.

\begin{table}[H]
\centering
\caption{Required sample size for 80\% power ($\alpha=0.05$, two-sided) for $\Delta_{\LHR}$ under uncertain discriminatory performance and CVD incidence. $N_{\eqref{eq:N-formula}}$ is the uncorrected Wald size from Equation~\eqref{eq:N-formula}; $N_{\mathrm{final}}$ applies the safeguarded rule \eqref{eq:n_final} to the same expression, adding the continuity correction on $(p_1,p_0)$ and the events-per-parameter floor at $c=10$, so $N_{\mathrm{final}}\ge N_{\eqref{eq:N-formula}}$ by construction. The boldface row is the observed $\mathrm{CACS}\ge10$ operating point ($\Se=0.82$, $\Sp=0.68$, $\pi=0.072$). Entries with a superscript $\dagger$ are floored by the EPV rule $N_{\mathrm{EPV}}=\lceil 10/\pi\rceil$; in the remaining rows the Wald or continuity-corrected term is binding.}
\label{tab:cacs-sensitivity}
\small
\begin{tabular}{ccccccccc}
\toprule
\textbf{Se} & \textbf{Sp} & $\boldsymbol{\pi}$ & $\boldsymbol{p_1}$ & $\boldsymbol{p_0}$ & $\boldsymbol{\Delta_{\LHR}}$ & $\boldsymbol{\mathrm{HR}}$ & $\boldsymbol{N_{\eqref{eq:N-formula}}}$ & $\boldsymbol{N_{\mathrm{final}}}$ \\
\midrule
0.60 & 0.60 & 0.05 & 0.073 & 0.034 & 0.790 & 2.20 & 1049 & 1052\\
0.60 & 0.60 & 0.07 & 0.101 & 0.048 & 0.782 & 2.18 & 766 & 769\\
0.60 & 0.60 & 0.10 & 0.143 & 0.069 & 0.769 & 2.16 & 554 & 558\\
0.70 & 0.70 & 0.07 & 0.149 & 0.031 & 1.629 & 5.10 & 202 & 203\\
0.80 & 0.60 & 0.07 & 0.131 & 0.024 & 1.733 & 5.66 & 234 & 237\\
0.60 & 0.80 & 0.07 & 0.184 & 0.036 & 1.707 & 5.51 & 161 & 161\\
0.80 & 0.70 & 0.05 & 0.123 & 0.015 & 2.175 & 8.80 & 208 & 208\\
\textbf{0.82} & \textbf{0.68} & \textbf{0.072} & \textbf{0.166} & \textbf{0.020} & \textbf{2.188} & \textbf{8.92} & \textbf{155} & \textbf{155}\\
0.80 & 0.70 & 0.07 & 0.167 & 0.021 & 2.151 & 8.60 & 152 & 152\\
0.80 & 0.70 & 0.10 & 0.229 & 0.031 & 2.117 & 8.30 & 110 & 110\\
0.85 & 0.80 & 0.07 & 0.242 & 0.014 & 2.986 & 19.80 & 99 & 143$^{\dagger}$\\
0.80 & 0.85 & 0.07 & 0.286 & 0.017 & 2.956 & 19.22 & 81 & 143$^{\dagger}$\\
0.90 & 0.90 & 0.07 & 0.404 & 0.008 & 4.129 & 62.11 & 74 & 143$^{\dagger}$\\
\bottomrule
\end{tabular}
\end{table}

Investigators can identify dominant drivers of required $N$ during routine protocol drafting rather than after a separate simulation cycle, and can set design conservatism by selecting the least favorable plausible combination or the boundary of a joint confidence region for $(\Se,\Sp)$ from pilot data.

\section{Discussion}
\label{sec:discussion}

Predictive model developers and clinical investigators routinely characterize test performance by sensitivity and specificity, yet statistical power for a validation study is typically calculated on regression-effect scales that require separate parameterization. We bridge this gap by mapping $(\Se,\Sp,\pi)$ directly to Wald variance and required $N$ for the risk contrast of interest. The formulas are closed-form, so an investigator can evaluate power at several $(\Se,\Sp,\pi)$ inputs in seconds without a simulation loop, which requires extensive time and computing resources.

Se/Sp planning and effect-size planning are the same design problem in different parameterizations. Working in Fisher information shows how prevalence and threshold act on the contrast scale. The predictive-value literature and the regression-power literature have long coexisted with limited interactions. Classic literature established that PPV and NPV depend strongly on prevalence \cite{brenner_variation_1997,murphy_relationship_1983,pepe_integrating_2008}, while others, working from the other direction, developed power formulas parameterized by regression coefficients and event counts \cite{shieh_power_2005,schoenfeld_sample-size_1983}. Steinberg et al.~\cite{steinberg_sample_2009} came closest to connecting the two by providing PPV/NPV precision formulas for case--control designs, where the two predictive values are correlated through $(\widehat{\Se},\widehat{\Sp})$, but they stopped at predictive-value precision rather than carrying the calculation through to risk contrasts. Concretely, once $(\Se,\Sp,\pi)$ are specified, the stratum risks $(p_1,p_0)$ are already implicit in Steinberg et al.'s setup. For example, a test with $\Se=0.85$ and $\Sp=0.90$ at $\pi=0.20$ immediately fixes both predictive values and the implied risk difference between strata. From there, the delta method delivers variance, power, and sample size on whichever contrast scale (e.g., $\RD$, $\LRR$, $\LOR$, or fixed-horizon $\LHR$) the investigator needs. Beyond assembling these components into a structured design workflow (Algorithm~1), we want to highlight the correction factor $K$ (Appendix~\ref{app:schoenfeld}), which restores the two approximations the Schoenfeld event-based formula makes---small event probabilities and comparable stratum risks. In the CACS example the second is what matters because $K=1.55$ is almost entirely a stratum-imbalance correction.

Variance depends on the separation $f(p_1)-f(p_0)$, and the stratum balance $\tau=\Pr(X=1)$. Extreme prevalence or thresholds inflate variance through $\tau$, regardless of discrimination. For instance, a biomarker with $\Se=0.85$ and $\Sp=0.90$ evaluated in a population with 10\% prevalence rather than 30\% requires about twice the uncorrected sample size on the log-contrast scales ($\LRR$, $\LOR$, $\LHR$) and about five times as many on the risk-difference scale, all else being equal; after the events-per-parameter floor is applied the ratio is close to three-fold on every scale, because at that point the floor rather than the variance sets the size. The C-optimal surfaces illustrate how required sample size can shift by multiples under modest changes in operating characteristics.

The CACS case study illustrates where this sensitivity is practically consequential. At the observed operating point the implied effect is large ($\mathrm{HR}\approx8.9$), yet the low CVD incidence keeps the effective event count modest, so the required $N$ for 80\% power is $N_{\mathrm{final}}=155$ rather than the few-dozen sample sizes seen with more common outcomes, and the design carries only about eleven expected events at that size. Under the lower-performance and lower-incidence scenarios examined in Section~\ref{sec:input-uncertainty}, attenuation of risk separation increases design requirements sharply, suggesting that modest degradation in $(\Se,\Sp)$ or a rarer outcome can move a design from feasible to impractical in biomarker validation settings.

\paragraph{Limitations.}
Power calculations rely on first-order, large-sample approximations, which are convenient analytically but inevitably imperfect in finite samples. Sparse data, extreme imbalance, or near-separation can lead to underestimated variance \cite{hauck1977wald,van_smeden_no_2016}. We incorporate continuity corrections and minimum expected-events constraints, which improve finite-sample behavior but do not replace simulation when severe separation or extreme imbalance is anticipated. In such settings, penalized methods should be prespecified. In our simulations, calibration was generally acceptable except in high-effect or sparse-cell regimes.

We also treat the cross-stratum covariance between $\widehat{p}_1$ and $\widehat{p}_0$ as negligible at first order. This term is exactly zero on the risk scale and $O(N^{-3})$ on the log scales, and was empirically negligible for moderate sample sizes (Appendix~\ref{app:fisher-info}), but in very small studies, it may contribute non-negligible approximation error.

In addition, the derivations are marginal in the test result $X$ and do not address covariate adjustment, interactions, or clustering. Extending the framework to these settings would require additional assumptions about covariate distributions and their associations with test performance and outcome. 

Finally, for time-to-event outcomes, incorporating dynamic predictive-value frameworks with time-varying PPV/NPV, together with design approaches for delayed or non-proportional effects \cite{xu_designing_2018, zheng_semiparametric_2010}, would allow predictive performance and effect size to evolve over follow-up and link prediction-oriented evaluation with time-dependent risk modeling.

\paragraph{Practical scope and positioning.}
The framework is most useful when $(\Se,\Sp,\pi)$ are the natural inputs---typically early validation and feasibility studies. In such settings, Algorithm~1 and the sensitivity table (Table~\ref{tab:cacs-sensitivity}) allow investigators to explore prevalence, threshold, and enrollment trade-offs without simulation. Final inference should still be conducted under the prespecified analysis model.

For large pivotal trials with well-established inputs and dedicated simulation infrastructure, regression-based or simulation-based planning may remain preferable. Likewise, in settings with extreme effects, severe imbalance, or anticipated separation, simulation under the intended analysis model (including penalization if prespecified) provides a more reliable assessment than first-order approximations alone.

Last but not least, the closed-form expressions also support inversion of the forward calculations to facilitate predictive model design. For a fixed enrollment and prevalence, one can characterize the set of $(\Se, \Sp)$ pairs sufficient to achieve target power, defining a minimum performance boundary in operating-characteristic space. This boundary clarifies whether a test's discriminatory capacity is adequate for a study of a given size. Its dependence on prevalence and effect-size scale follows directly from the expressions derived here.

\appendix
\setcounter{figure}{0}
\setcounter{table}{0}
\setcounter{equation}{0}
\renewcommand{\thefigure}{A\arabic{figure}}
\renewcommand{\thetable}{A\arabic{table}}
\renewcommand{\theequation}{A\arabic{equation}}
\renewcommand{\theHequation}{A\arabic{equation}}

\section{Appendix}
\label{sec:appendix}

\subsection{Fisher information and covariance of test-positive and test-negative strata}
\label{app:fisher-info}

The main text provides the analytic order argument for treating the cross-stratum covariance as negligible in first-order variance calculations. This appendix evaluates the closed-form delta-method covariance and reports finite-sample simulations to quantify its magnitude.

For the random stratum size, $N_+ \sim \mathrm{Bin}(N,\tau)$, so
\[
\Var(N_+) = N\tau(1-\tau)=O(N), \qquad N_+-\tau N=O_p(N^{1/2}).
\]
Using
\[
\frac{1}{N_+}
=
\frac{1}{\tau N}
-
\frac{N_+-\tau N}{\tau^2 N^2}
+
O_p(N^{-2}),
\]
the random-denominator term is $O_p(N^{-3/2})$. The same order holds for $1/N_-$. These terms affect the variance only, because $\mathbb{E}(\widehat{p}_x\mid N_x)=p_x$ does not depend on $N_x$, so $\Cov(\widehat{p}_1,\widehat{p}_0)=0$ for non-empty strata. On the log scales, $\Cov(1/N_+,1/N_-)\approx-\Var(N_+)/[\tau^2(1-\tau)^2N^4]=O(N^{-3})$, and with the $O(1/N_x)$ bias of $f(\widehat{p}_x)$ this gives $\Cov\{f(\widehat{p}_1),f(\widehat{p}_0)\}=O(N^{-3})$ against leading variance terms of $O(N^{-1})$.

To assess finite-sample magnitude, we simulated multinomial $2\times2$ tables over
\begin{align*}
    \mathrm{Se},\mathrm{Sp} &\in\{0.65, 0.70, 0.75, 0.80, 0.85, 0.90, 0.95\}, \\
    \pi &\in \{0.10,0.30,0.50,0.70\}, \\
    N &\in  \{50,100,200,400,800,1600,3200,6400\},
\end{align*}
with $R=10^5$ replicates per configuration; large $R$ was used to estimate very small covariances stably.

For each configuration we computed both the closed-form delta-method covariance of $(\widehat{\mathrm{PPV}},\widehat{\mathrm{NPV}})$ under the multinomial model and its Monte Carlo counterpart.

The closed-form value is zero to machine precision at every design point on the grid (largest absolute value $4\times10^{-19}$ at $N=50$, falling to $3\times10^{-21}$ at $N=6400$). This is the exact zero the argument above predicts rather than a numerical coincidence: $\widehat{\mathrm{PPV}}$ depends on the table only through $(n_{11},n_{10})$ and $\widehat{\mathrm{NPV}}$ only through $(n_{00},n_{01})$, and the multinomial cross-cell covariances $-Np_ip_j$ enter the two gradients with weights that sum to zero at first order.

\begin{figure}[t]
\centering
\includegraphics[width=1\textwidth]{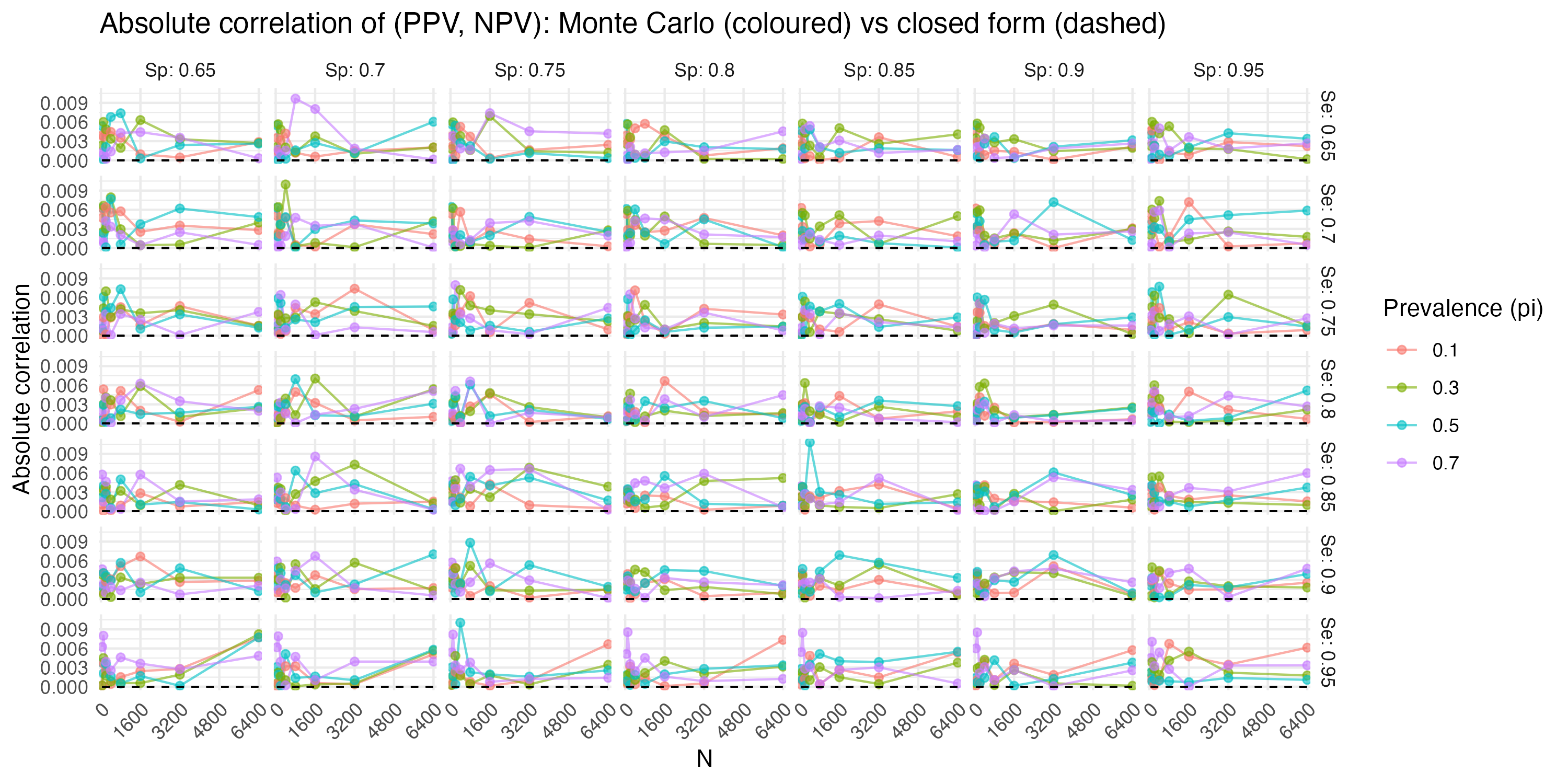}\\[4pt]
\includegraphics[width=1\textwidth]{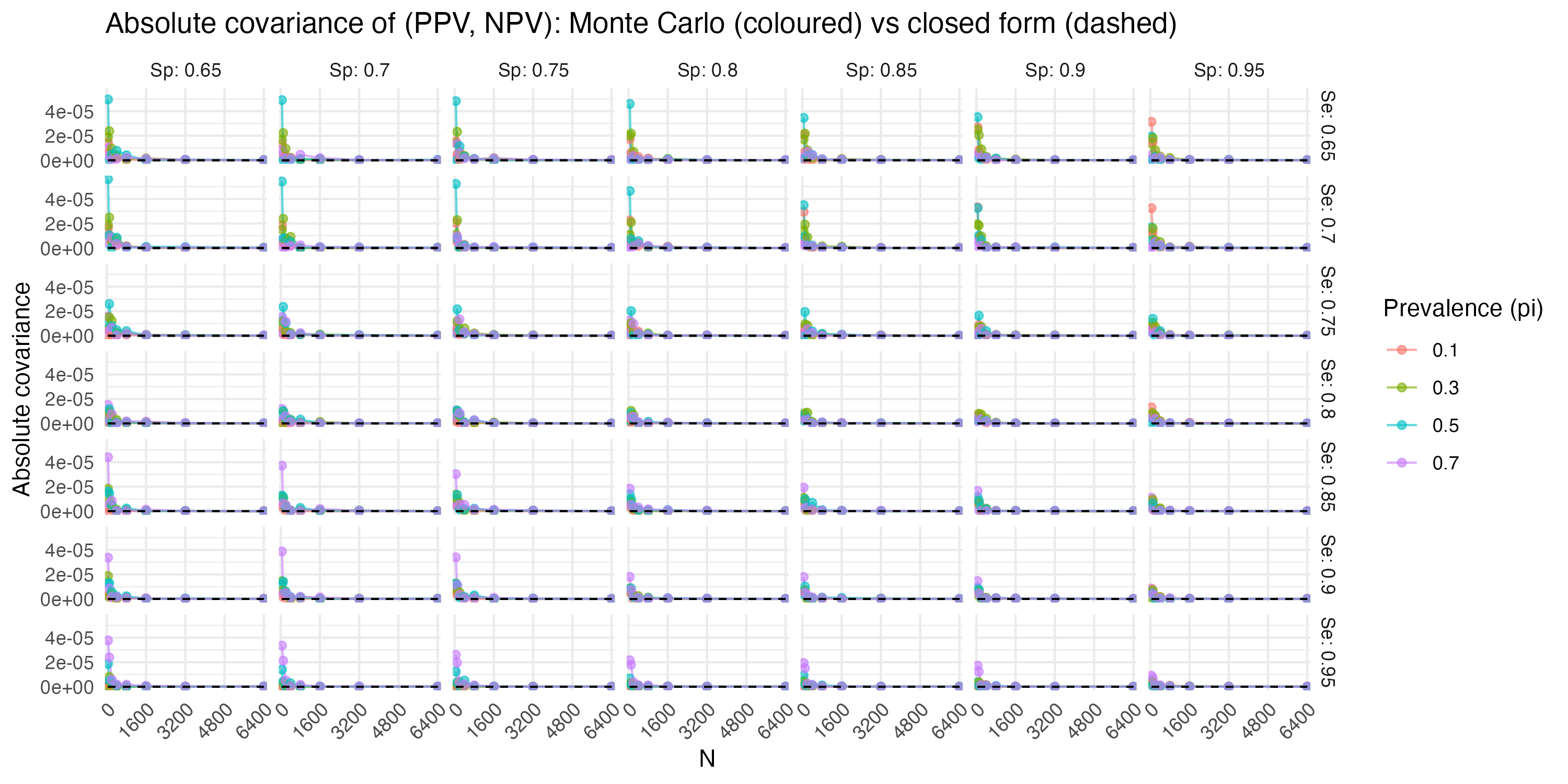}
\caption{Absolute values of correlation (top) and covariance (bottom) between estimated PPV and NPV across sample sizes and operating points. Coloured lines are Monte Carlo estimates by prevalence; the dashed black line is the closed-form delta-method value, which lies on the axis because it is zero to machine precision at every design point. The Monte Carlo values sit within a factor of about two of the simulation resolution limit at every $N$, so they bound the covariance as negligible without identifying its rate.}

\noindent\textit{Alt text: Two stacked line-and-point panels showing absolute correlation (top) and covariance (bottom) between estimated PPV and NPV across sample sizes, by prevalence; Monte Carlo values remain at the simulation resolution limit at all $N$, consistent with a closed-form value of exactly zero.}
\label{fig:abs-cov-corr}
\end{figure}

The Monte Carlo values are correspondingly small but are limited by simulation error rather than by the estimand. The Monte Carlo standard error of a covariance is approximately $\mathrm{sd}(\widehat{\mathrm{PPV}})\,\mathrm{sd}(\widehat{\mathrm{NPV}})/\sqrt{R}$, which itself decays like $N^{-1}$; across the grid the largest observed $|\Cov|$ stays within a factor of $1.8$--$2.5$ of that resolution limit at every $N$ (for example $5.6\times10^{-5}$ against a limit of $2.9\times10^{-5}$ at $N=50$, and $4.1\times10^{-7}$ against $2.2\times10^{-7}$ at $N=6400$). The empirical correlation behaves the same way, fluctuating without trend between $0.006$ and $0.011$ across all sample sizes. The simulation therefore confirms that the cross-stratum covariance is negligible at design-relevant sample sizes, but the Monte Carlo values reflect simulation noise rather than the estimand, which is why the conclusion rests on the analytic argument rather than on Figure~\ref{fig:abs-cov-corr}. Together these results support using the block-diagonal first-order approximation in Fisher-information and sample-size derivations. In settings with very small $N$ or extreme $\tau$, approximation error can be larger and should be checked.

\begin{figure}
    \centering
    \includegraphics[width=1\linewidth]{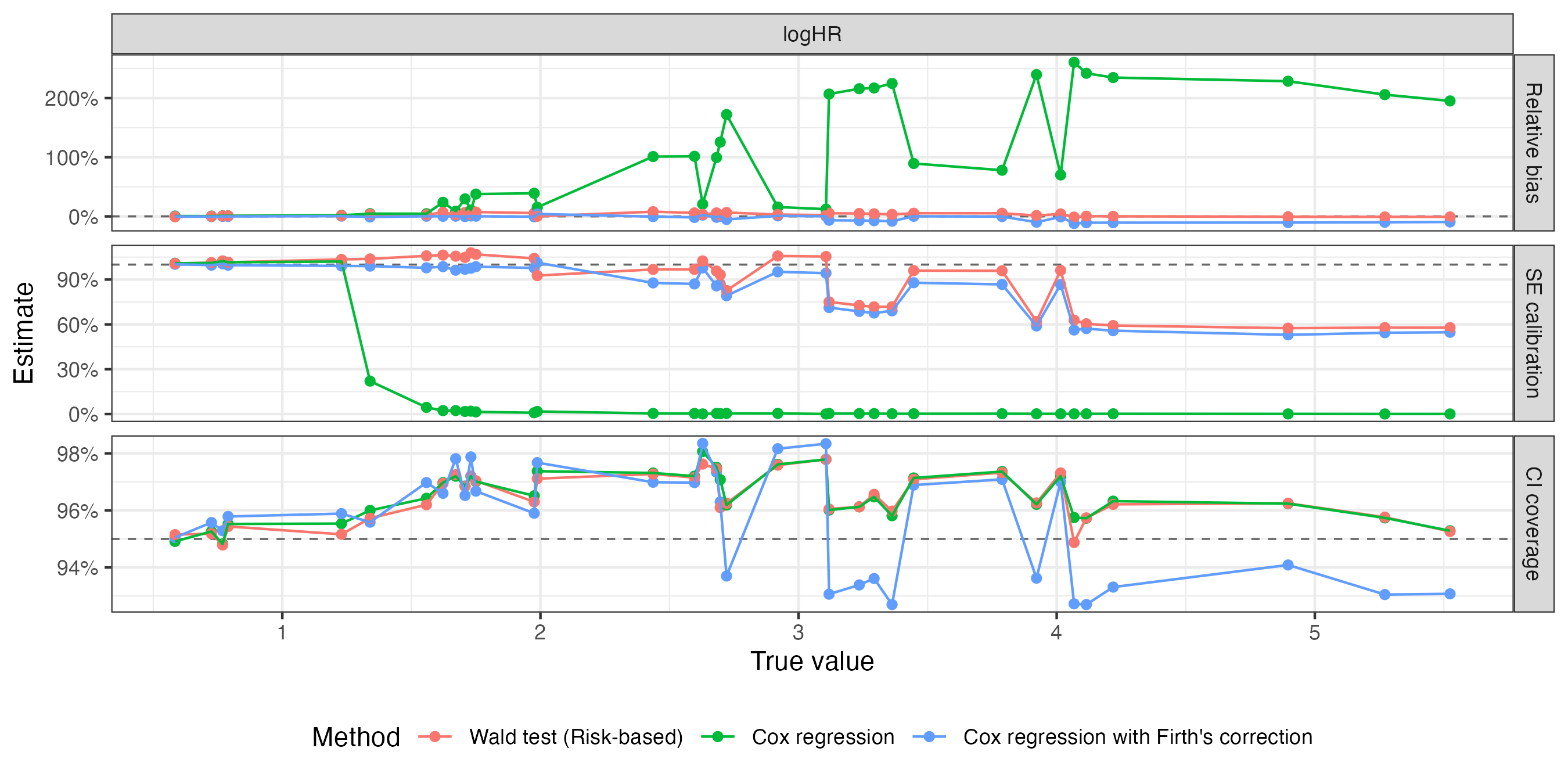}
    \caption{Relative bias, SE calibration, and 95\% CI coverage for the fixed-horizon logHR, comparing the risk-based Wald estimator with Cox partial-likelihood estimation with and without Firth correction. The uncorrected Cox estimator's calibration ratio collapses towards zero beyond $\LHR\approx1.3$: its model-based standard error inflates far faster than its sampling variability, so its intervals stay wide (coverage at or above nominal) while its Wald test loses power.}

    \noindent\textit{Alt text: Relative bias, SE calibration, and 95\% CI coverage for the fixed-horizon logHR comparing the risk-based Wald estimator against uncorrected and Firth-corrected Cox models; the uncorrected Cox calibration ratio collapses toward zero at large logHR while the other two estimators remain close to their reference lines.}
    \label{fig:app_bias_plot}
\end{figure}

\subsection{From the Wald-type Variance to Schoenfeld's Formula}
\label{app:schoenfeld}

This appendix derives the Schoenfeld event-based form from the Fisher-information variance for the $\LHR$ contrast in the main text. At the design stage, $\tau$, $p_1$, $p_0$, and $t^\ast$ are treated as fixed inputs, and variance expressions are first-order large-sample plug-in approximations at those values.

\paragraph{Wald-type variance}
From Section~\ref{sec:var-risk}, any smooth effect size $g(p_1,p_0)$ satisfies
\begin{equation}
\Var(\widehat{\Delta})
\approx
\frac{1}{N}
\left[
c_1^2\,\frac{p_1(1-p_1)}{\tau}
+
c_0^2\,\frac{p_0(1-p_0)}{1-\tau}
\right],
\qquad
p_{\mathrm{eff}}=\tau p_1+(1-\tau)p_0,
\label{eq:app-general}
\end{equation}
where $c_x=\partial g/\partial p_x$ and $D=N p_{\mathrm{eff}}$ is the expected number of events.

For the log-hazard ratio,
\[
\Delta_{\LHR}
= \log[-\log(1-p_1)] - \log[-\log(1-p_0)],
\]
so
\begin{equation}
c_1=\frac{1}{(1-p_1)[-\log(1-p_1)]},
\qquad
c_0=-\frac{1}{(1-p_0)[-\log(1-p_0)]}.
\label{eq:app-derivs}
\end{equation}

Substituting \eqref{eq:app-derivs} into \eqref{eq:app-general} yields the Fisher-information variance \emph{without the small-risk approximation}:
\begin{equation}
\Var(\widehat{\Delta}_{\LHR})
\approx
\frac{1}{N}
\left[
\frac{1}{\tau}
\frac{p_1}{(1-p_1)[-\log(1-p_1)]^2}
+
\frac{1}{1-\tau}
\frac{p_0}{(1-p_0)[-\log(1-p_0)]^2}
\right].
\label{eq:app-exact}
\end{equation}

\paragraph{Small-to-moderate risk approximation and reduction to Schoenfeld's formula.}
When risks are not extreme,
\[
(1-p_x)\approx 1, 
\qquad
-\log(1-p_x)\approx p_x,
\]
so the exact terms simplify:
\[
\frac{p_x}{(1-p_x)[-\log(1-p_x)]^2}
\approx \frac{1}{p_x}.
\]
Thus \eqref{eq:app-exact} becomes
\begin{equation}
\Var(\widehat{\Delta}_{\LHR})
\approx
\frac{1}{N}
\left(
\frac{1}{\tau p_1}
+
\frac{1}{(1-\tau)p_0}
\right).
\label{eq:app-mid}
\end{equation}

The Schoenfeld reduction requires two approximations: the small-to-moderate risk approximation above and an additional balance condition $p_1\approx p_0\approx p_{\mathrm{eff}}$ (i.e., no strong group-risk imbalance).
If group-specific risks are similar ($p_1\approx p_0\approx p_{\mathrm{eff}}$), then
\[
\frac{1}{\tau p_1}
+\frac{1}{(1-\tau)p_0}
\approx
\frac{1}{p_{\mathrm{eff}}\tau(1-\tau)}.
\]
Using $D=N p_{\mathrm{eff}}$,
\begin{equation}
\Var(\widehat{\Delta}_{\LHR})
\approx
\frac{1}{D\,\tau(1-\tau)},
\label{eq:app-schoenfeld}
\end{equation}
which is the classic Schoenfeld event-based variance.

\paragraph{Correction when either approximation fails.}
Two approximations were used above: $-\log(1-p_x)\approx p_x$, which fails when event probabilities are moderate or large, and $p_1\approx p_0\approx p_{\mathrm{eff}}$, which fails when the two predictive strata separate. In either case, retain the exact variance \eqref{eq:app-exact}, or equivalently factor it as
\begin{equation}
\Var(\widehat{\Delta}_{\LHR})
\approx
\frac{K(p_1,p_0,\tau)}{D\,\tau(1-\tau)},
\label{eq:app-correction}
\end{equation}
where the correction factor is
\[
K(p_1,p_0,\tau)
=
p_{\mathrm{eff}}
\left[
(1-\tau)\frac{p_1}{(1-p_1)[-\log(1-p_1)]^2}
+
\tau\frac{p_0}{(1-p_0)[-\log(1-p_0)]^2}
\right].
\]
By construction $K$ makes \eqref{eq:app-correction} agree with the first-order Fisher-information expression exactly, and it absorbs both approximations at once. $K\approx1$ requires small risks \emph{and} $p_1\approx p_0\approx p_{\mathrm{eff}}$: substituting $p_1=p_0=p$ in the small-risk limit $K\approx p_{\mathrm{eff}}[(1-\tau)/p_1+\tau/p_0]$ returns $K=1$ identically, whereas any separation between $p_1$ and $p_0$ inflates $K$ even when both risks are small. The two contributions can be read off by comparing $K$ with its small-risk limit: at the CACS operating point of Section~\ref{sec:casestudy} ($p_1\approx0.17$, $p_0\approx0.02$, $\tau\approx0.36$), $K=1.554$ while the small-risk limit is $1.553$, so the risk-magnitude term contributes a factor of $1.001$ and stratum imbalance contributes the remaining $1.553$. Note also that $K$ equals, up to the factor $\pi\,h(\pi)\approx1$, the ratio of the per-subject variance at the design point to its value when both stratum risks are set to $\pi$; this is why $K$ and that ratio take the same numerical value in Section~\ref{sec:casestudy}.

\subsection{Optimality criteria and efficiency metrics}
\label{app:d-opt}

For statistical efficiency using C-optimality (single estimand), $\Eff=\Delta^2/v$ as in Equation~\eqref{eq:E1}, so maximizing $\Eff$ and minimizing the per-subject variance \eqref{eq:v-persubject} for fixed $\Delta$ are the same problem.

For D-optimality (joint estimand), with $\bphi=(\mathrm{PPV},\mathrm{NPV})^\top$,
\[
\max_{\mathcal D}\det\{\mathcal I_{\bphi}\}
\quad\Longleftrightarrow\quad
\min_{\mathcal D}\det\{\mathcal I_{\bphi}^{-1}\}.
\]
Moreover, $\det(\mathcal I_{\bphi}^{-1})^{1/2}$ is proportional to the area (volume in higher dimension) of the asymptotic confidence ellipsoid, so D-optimality targets joint uncertainty.

\begin{figure}[ht]
\centering
\includegraphics[width=\linewidth]{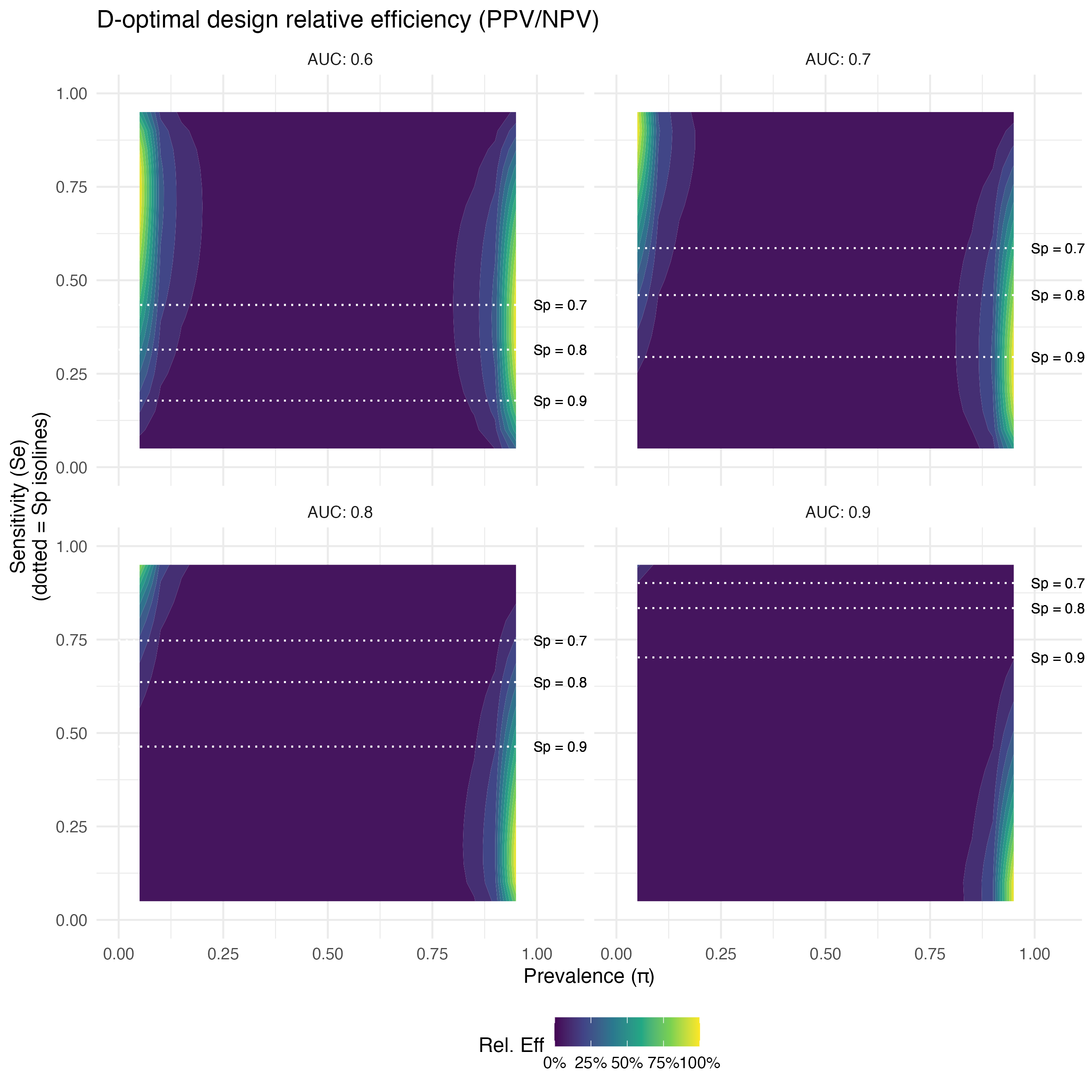}
\caption{D-optimal design efficiency surface.}

\noindent\textit{Alt text: Heatmap of D-optimal design efficiency over prevalence and sensitivity; efficiency is highest near the extremes of prevalence or sensitivity, the opposite pattern from the C-optimal surfaces.}
\label{fig:dopt-surface}
\end{figure}

Because Bernoulli variance terms shrink near 0 or 1, D-optimality can favor operating points where one predictive value approaches an extreme---often clinically unrealistic. We therefore use D-optimality as a diagnostic for joint $(\mathrm{PPV},\mathrm{NPV})$ uncertainty rather than as the primary design criterion for effect-size studies (Figure~\ref{fig:dopt-surface}).

\begin{figure}[ht]
\centering
\includegraphics[width=\linewidth]{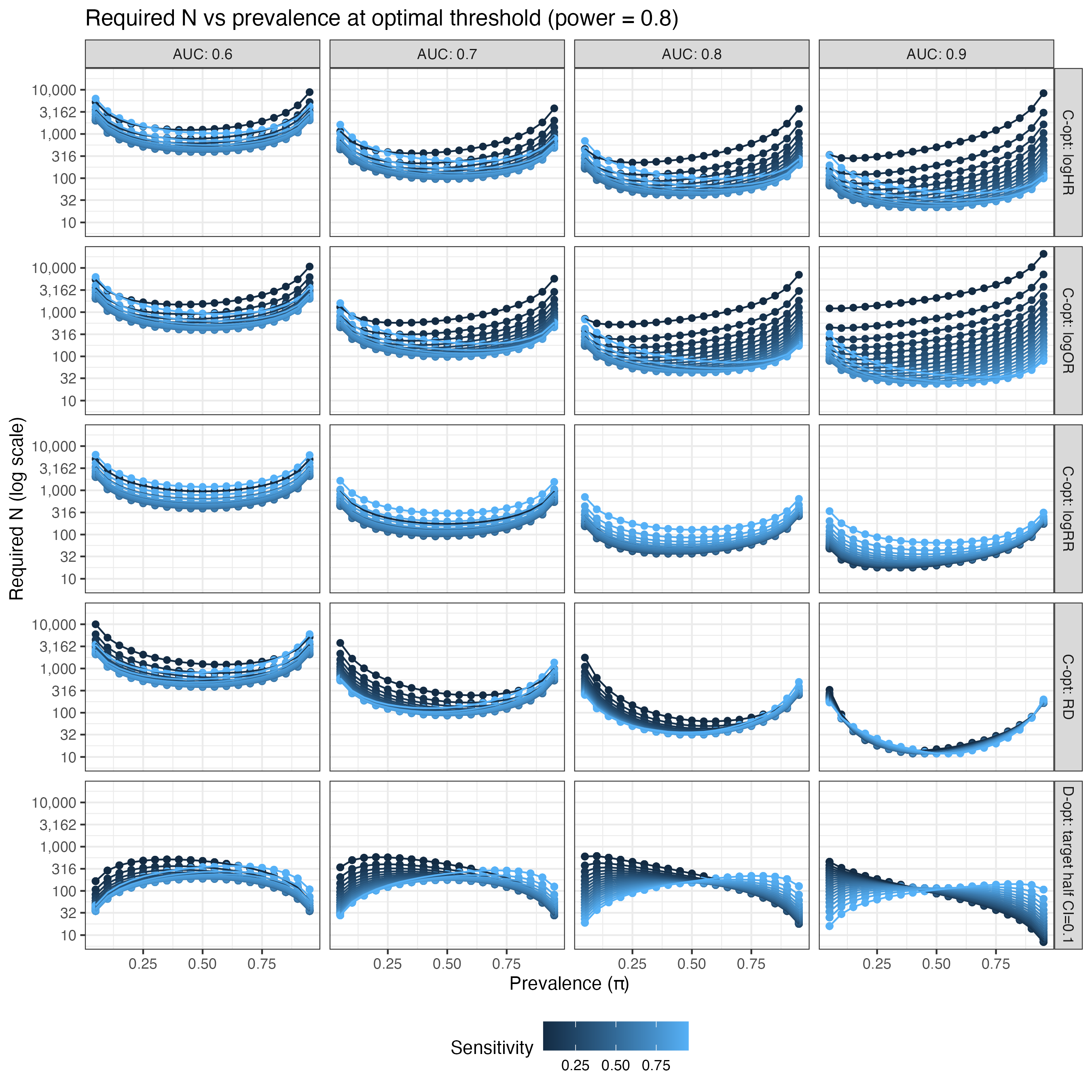}
\caption{Relations between statistical efficiency and sensitivity/specificity, prevalence, and sample size.}

\noindent\textit{Alt text: Grid of required-sample-size curves over prevalence, split by AUC, estimand, and design criterion (C-optimal vs.\ D-optimal); C-optimal required $N$ is U-shaped in prevalence while D-optimal required $N$ peaks at intermediate prevalence, the reverse pattern.}
\label{fig:cdopt-n}
\end{figure}

Figure~\ref{fig:cdopt-n} compares required sample size under the two criteria across prevalence, AUC, estimands, and threshold-induced operating points. For the C-optimal rows, required $N$ is U-shaped in prevalence with a minimum at intermediate values, and the whole family shifts downward as AUC increases---the information structure of Equation~\eqref{eq:vardelta-final}, in which both predictive strata must be populated for the contrast to be estimable.

The D-optimal row runs the other way: required $N$ peaks at intermediate prevalence and falls towards both extremes. This is the boundary artifact of Section~\ref{subsec:dopt} seen on the sample-size scale. Precision in $(\mathrm{PPV},\mathrm{NPV})$ is easiest to buy where one predictive value is pushed against $0$ or $1$ and its Bernoulli variance vanishes, which is exactly where the risk contrast is least estimable. The two criteria therefore recommend opposite operating points, reinforcing our use of D-optimality as a diagnostic for predictive-value precision rather than as a design criterion for effect-size studies. Sizes on the D-optimal row are those needed for a $95\%$ interval half-width of $0.1$, expressed as the geometric mean of the PPV and NPV standard errors; since $|\mathcal{I}_{\bphi}|$ grows like $N^2$, the required $N$ scales with the square root of the per-subject variance product.

\bibliographystyle{chicago}
\bibliography{references}

\end{document}